\documentclass[a4paper,twocolumn,longbibliography,accepted=2022-03-16]{quantumarticle}
\pdfoutput=1
\usepackage[numbers,sort&compress]{natbib}
\usepackage{graphicx,times}
\usepackage{color}

\usepackage{sansmath}
\usepackage{amssymb}

\usepackage{subeqnarray}
\usepackage{diagbox}
\usepackage{makecell}
\usepackage{amsmath}
\usepackage{mathtools}
\usepackage{empheq}
\usepackage[all]{xy}

\newcommand{\msf}[1]{\mathsf{#1}}
\newcommand{\mcl}[1]{\mathcal{#1}}
\newcommand{\bsb}[1]{\boldsymbol{#1}}

\newcommand{\mathbbmm}[1]{\text{\usefont{U}{bbm}{m}{n}#1}}

\DeclareSymbolFont{greekletters}{OML}{FiraSans}{m}{n}

\DeclareMathOperator{\mUpsilon}{\msf{\Upsilon}}
\DeclareMathOperator{\mSigma}{\msf{\Sigma}}
\DeclareMathOperator{\mDelta}{\msf{\Delta}}
\DeclareMathOperator{\mcd}{\msf{c}_{\delta}}
\DeclareMathOperator{\mld}{\msf{l}_{\delta}}
\DeclareMathOperator{\mA}{\msf{A}}

\DeclareMathOperator{\mC}{\msf{C}}
\DeclareMathOperator{\mY}{\msf{Y}}
\DeclareMathOperator{\mZ}{\msf{Z}}
\DeclareMathOperator{\mS}{\msf{S}}
\DeclareMathOperator{\mP}{\msf{P}}
\DeclareMathOperator{\mQ}{\msf{Q}}

\DeclareMathOperator{\bU}{\bsb{U}}
\DeclareMathOperator{\bV}{\bsb{V}}
\DeclareMathOperator{\bW}{\bsb{W}}
\DeclareMathOperator{\bu}{\bsb{u}}

\DeclareMathOperator{\bPhi}{\bsb{\Phi}}
\DeclareMathOperator{\bQ}{\bsb{Q}}
\DeclareMathOperator{\bPi}{\bsb{\Pi}}

\DeclareMathOperator{\bX}{\bsb{X}}

\DeclareMathOperator{\be}{\bsb{e}}
\DeclareMathOperator{\br}{\bsb{r}}

\DeclareMathOperator{\mL}{\mcl{L}}
\DeclareMathOperator{\mT}{\mcl{T}}

\DeclareMathOperator{\mD}{\mcl{D}}

\DeclareMathOperator{\mW}{\mathcal{W}}

\DeclareMathOperator{\mt}{\msf{t}}

\DeclareMathOperator{\mcM}{\mcl{M}}
\DeclareMathOperator{\mM}{\msf{M}}

\DeclareMathOperator{\msD}{\msf{D}}

\DeclareMathOperator{\msF}{\msf{F}}
\DeclareMathOperator{\msG}{\msf{G}}
\DeclareMathOperator{\msJ}{\msf{J}}

\DeclareMathOperator{\bE}{\bsb{E}}

\newcommand{\ome}{{\omega\epsilon}}
\newcommand{\omel}{{\omega\lambda}}

\newcommand{\omep}{{\omega^{\prime}\epsilon^{\prime}}}

\newcommand{\mR}{\mathbbmm{R}}

\newcommand{\mone}{\mathbbmm{1}}

\begin{document}

\title{Canonical quantisation of telegrapher's equations
	coupled by ideal nonreciprocal elements}

\author{A. Parra-Rodriguez}
\affiliation{Department of Physics, University of the Basque Country UPV/EHU, Apartado 644, 48080 Bilbao, Spain}
\email{adrian.parra.rodriguez@gmail.com}
\orcid{0000-0002-0896-9452}
\author{I. L. Egusquiza}
\affiliation{Department of Physics, University of the Basque Country UPV/EHU, Apartado 644, 48080 Bilbao, Spain}
\orcid{0000-0002-5827-8027}

\begin{abstract}
We develop a systematic procedure to  quantise canonically Hamiltonians of light-matter models of transmission lines coupled through lumped linear lossless ideal nonreciprocal elements, that break time-reversal symmetry, in a circuit QED set-up. This is achieved through a description of the distributed subsystems in terms of both flux and charge fields. We prove that this apparent redundancy is required for the general derivation of the Hamiltonian for a wider class of networks. By making use of the electromagnetic duality symmetry in transmission lines (waveguides), we provide unambiguous identification of the physical degrees of freedom, separating out the nondynamical parts. This doubled description can also treat the case of other extended lumped interactions in a regular manner that presents no spurious divergences, as we show explicitly in the example of a circulator connected to a Josephson junction through a transmission line. This theory enhances the quantum engineering toolbox to design complex networks with nonreciprocal elements.
\end{abstract}

\maketitle
\section{Introduction}
The most sophisticated quantum processors available to date rely on cryogenic superconducting circuits \cite{Devoret:2013}, to the point that  on such devices  quantum algorithms have recently overtaken classical machines at simulating specific quantum evolutions \cite{Arute:2019}. Superconducting quantum circuits are based on microwave technology and typically modeled with lumped element capacitors, inductors, or Josephson junctions, as well as distributed elements in the form of transmission lines. Other devices  such as  superconducting 3D cavities are also used, and are included in minimal theoretical models via lumped-element (Foster) expansions of the low-lying frequency modes after a classical multi-physics analysis \cite{Nigg:2012,Solgun:2015}.

More recently, the miniaturisation of nonreciprocal (NR) devices, including the circulator or the gyrator \cite{Sliwa:2015,Chapman:2017,Mueller:2018,Kerckhoff:2015,Viola:2014,Mahoney:2017,Barzanjeh:2017}, that effectively break time-reversal symmetry (TRS) and can be used for noise isolation or signal routing \cite{Tellegen:1948}, has allowed their inclusion in the cryogenic chip. Thus the need arises for these final linear passive lossless elements to be included in the quantum circuit analysis \cite{YurkeDenker:1984,Chakravarty:1986,Yurke:1987,Werner:1991,Devoret:1997,Paladino:2003,Burkard:2004,Burkard:2005,Bergenfeldt:2012,Bourassa:2012,Nigg:2012,Bamba:2014,Solgun:2014,Solgun:2015,Mortensen:2016,Malekakhlagh:2016,Malekakhlagh:2017,Gely:2017,ParraRodriguez:2018,ParraRodriguez:2019,Minev:2020,Mariantoni:2020,Minev:2021,Rymarz:2018,Rymarz:2021,Egusquiza:2022} on equal footing with capacitors, inductors, and transmission lines, as they now present quantum coherence properties. Remember that in an electrical network, reciprocity is equivalent to the invariance of the system response under inversion of source and detector, and nonreciprocity is the lack of such invariance.

In this article we propose a systematic procedure to  quantise canonically Hamiltonians of light-matter models  of transmission lines pointwise coupled through generic linear ideal circulator systems, i.e., coupled through a lumped nonreciprocal element localised at a point in space. In other words, we consider nonreciprocal coupling between extended (one-dimensional waveguides) elements, which can be additionally coupled to other lumped networks (containing capacitors, inductors, Josephson junctions, etc.). 

This theory builds on and goes beyond previous works on canonical quantisation in circuit QED~\cite{Devoret:1997,YurkeDenker:1984,Paladino:2003,Burkard:2004,Burkard:2005,Ulrich:2016} with couplings between extended and lumped elements~ \cite{ParraRodriguez:2018,Bamba:2014,Bergenfeldt:2012,Mortensen:2016,Malekakhlagh:2016,Malekakhlagh:2017,Minev:2021}, and lumped-element nonreciprocal networks  \cite{ParraRodriguez:2019,Rymarz:2018,Rymarz:2021,Egusquiza:2022}. Because nonreciprocal elements introduce intrinsic constraints between charges and fluxes, we  describe the circuits under consideration systematically in terms of \emph{both} flux and charge variables \cite{Jeltsema:2009,Ulrich:2016}. We make essential use of electromagnetic duality to show that the apparent redundancy is not a hindrance, and construct an explicit Hamiltonian. Furthermore, when coupled to other circuit lumped elements this construction maintains  good ultraviolet behaviour  \cite{Malekakhlagh:2017,ParraRodriguez:2018}, as we exemplify at the end with a circuit containing a Josephson junction coupled through a transmission line to a circulator. This theory enhances the quantum engineering toolbox to analyse and design complex superconducting circuits based on nonreciprocal elements. 

\section{Transmission lines in the doubled space}
\begin{figure}[h!]
	\centering
	\includegraphics[width=0.8\linewidth]{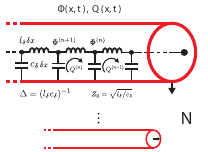}
	\caption{Differential model of a transmission line with primary parameters, i.e.,  inductance ($l_\delta$) and capacitance ($c_\delta$) per unit length, or secondary parameters, i.e.,  phase velocity ($\Delta$) and characteristic impedance ($Z_0$). Node-flux $\Phi_i(t)$ and loop-charge $Q_i(t)$ variables are a redundant set of degrees of freedom for this system. In this article we work with the flux $\Phi(x,t)$ and charge $Q(x,t)$ fields, i.e., in the continuum limit $\delta x\to0$. We consider $N$ such lines in the text.}
	\label{fig:TL_ds}
\end{figure}

Transmission lines (TLs) are physical media that confine electromagnetic fields, effectively reducing the number of relevant dimensions to one \cite{Pozar:2009}. In particular, Maxwell's equations for transversal-electromagnetic modes supported inside of $N$ TLs simplify to distributed Kirchhoff's equations, commonly known as {\it telegrapher's equations} \cite{Pozar:2009}, which in the continuum limit are expressed in column vectors of flux $\bPhi(x,t)$ and charge fields $\bQ(x,t)$ as 
\begin{eqnarray}
	\mcd\ddot{\bPhi}(x,t)&=&\dot{\bQ}'(x,t),\quad
	\mld\ddot{\bQ}(x,t)=\dot{\bPhi}'(x,t).\label{eq:telegraphers}
\end{eqnarray}
We use boldface to denote column vectors, in this case with $N$ components. Here, $\mcd$, $\mld$ are macroscopic capacitance and inductance per-unit-length diagonal matrices describing the $N$ lines, and we use the standard notation $\dot{f}=\partial_t f$ and $f'=\partial_x f$. In  what follows the time and space variables will be implicit, except where required for clarity or emphasis. The flux and charge real fields are related to the voltage (electric) and current (magnetic) fields via $\dot{\bPhi}=(\dot{\Phi}_1, \dot{\Phi}_2, ..., \dot{\Phi}_N)^T$, and $\dot{\bQ}=(\dot{Q}_1, \dot{Q}_2, ..., \dot{Q}_N)^T$ respectively, see Fig. (\ref{fig:TL_ds}).

For the sake of clarity, we will work through the rest of the article with rescaled flux and charge fields,  $\sqrt{\mcd}\bPhi\rightarrow\bPhi$ and $\bQ/\sqrt{\mcd}\rightarrow\bQ$. Then Eqs. (\ref{eq:telegraphers}) are the
Euler-Lagrange (E-L) equations for the {\it telegrapher's} Lagrangian \cite{Jeltsema:2009}
\begin{equation}
	L_{\mathrm{TG}}= \int_{\mcl{I}}dx\,\frac{1}{2}\left[\dot{\bPhi}^T\dot{\bPhi} +\dot{\bQ}^T\mDelta^{-1}\dot{\bQ}-\dot{\bQ}^T\bPhi'-\dot{\bPhi}^T\bQ'\right]\label{eq:TL_Lag_ds},
\end{equation}
where $\mDelta=\msf{c}_\delta^{-1/2}\msf{l}_\delta^{-1}\msf{c}_\delta^{-1/2}$ is the  velocity matrix, and $\mcl{I}$ is the interval of the TL, either finite or the half-line. Notice that $\bPhi$ and $\bQ$ are \emph{not} conjugate variables  here. Both of them are coordinates in the configuration space. 

This description has a gauge symmetry, in that Eqs. \eqref{eq:telegraphers} are invariant under the transformations $\bPhi(x,t)\to\bPhi(x,t)+\mathbf{A}(x)$, $\bQ(x,t)\to\bQ(x,t)+\mathbf{B}(x)$. The Lagrangian $L_{\mathrm{TG}}$ is not invariant under these transformations, but it transforms simply by the addition of a total time derivative. Thus, by a gauge choice, one can work with either just flux or just charge fields. If only flux, then charge counterparts will appear, now yes, as the conjugate momenta in a Hamiltonian. With the proper gauge choice, in fact, Eqs. (\ref{eq:telegraphers}) become the wave equations
\begin{eqnarray}
	\ddot{\bPhi}(x,t)-\mDelta\bPhi''(x,t)&=&0,\label{eq:wave_Phi}\\
	\ddot{\bQ}(x,t)-\mDelta\bQ''(x,t)&=&0,\label{eq:wave_Q}
\end{eqnarray}
as derived from the more commonly used Lagrangians in the superconducting quantum technologies community \cite{YurkeDenker:1984,Chakravarty:1986,Yurke:1987,Werner:1991,Devoret:1997,Paladino:2003,Blais:2004,Houck:2008,Bourassa:2009,Clerk:2010,Koch:2010,Filipp:2011,Bergenfeldt:2012,Bourassa:2012,Peropadre:2013,Sundaresan:2015,Malekakhlagh:2016,Mortensen:2016,Roy:2016,Vool:2017,Roy:2018,ParraRodriguez:2018,Malekakhlagh:2017} 
\begin{eqnarray}
	L_{\bPhi}&=&\int_{\mcl{I}}dx\,\frac{1}{2}\left[\dot{\bPhi}^T\dot{\bPhi}-(\bPhi')^T\mDelta\bPhi'\right],\label{eq:TL_Lag_Phi}\\
	L_{\bQ}&=&\int_{\mcl{I}}dx\,\frac{1}{2}\left[\dot{\bQ}^T\mDelta^{-1}\dot{\bQ}-(\bQ')^T\bQ'\right].\label{eq:TL_Lag_Q}
\end{eqnarray}
It is worth remarking here that the three Lagrangians above present physical time-reversal symmetry, see Appendix~\ref{sec:time-reversal}.

The fact that the telegrapher's Lagrangian presents this gauge invariance can be stated in a different way by saying that it includes nondynamical degrees of freedom, those corresponding to the $\mathbf{A}(x)$ and $\mathbf{B}(x)$ functions, for which there is no evolution. Those redundant variables are discarded in configuration space in this case by moving from $L_{\mathrm{TG}}$ to $L_{\bPhi}$ or $L_{\bQ}$, leaving a residual invariance under displacement by a global (homogeneous) constant.

  Notice, furthermore, that $L_{\bPhi}$ and $L_{\bQ}$ are mapped into each other as $L_{\bPhi}\leftrightarrows-L_{\bQ}$ by \emph{duality}, with the mappings
  \begin{subequations}
  \begin{align}
    \label{eq:dualityblock}
    \dot{\bPhi}&\leftrightarrows\bQ'\,, \\
    \mathsf{\Delta}\bPhi'&\leftrightarrows\dot{\bQ}.
  \end{align}
\end{subequations}
They are equivalent, in that a solution of the E-L equations for one of them becomes, under this duality operation, a solution of the E-L equations for the dual one. Explicitly, if $\bPhi(x,t)$ is a solution of Eq. \eqref{eq:wave_Phi}, then
  \begin{equation}
    \label{eq:solduality}
    \bQ(x,t):=\int^xd\chi\,\dot{\bPhi}(\chi,t)
  \end{equation}
  is a solution of Eq. \eqref{eq:wave_Q}. This equivalence does not however mean that both viewpoints will be equally useful in other contexts.

In fact, although the earliest works \cite{YurkeDenker:1984,Chakravarty:1986,Yurke:1987,Werner:1991} used the charge field Lagrangian (\ref{eq:TL_Lag_Q}), problems arose in the addition of nonlinear elements. For instance, Yurke and Denker \cite{YurkeDenker:1984} did not derive a Hamiltonian,  and Yurke  \cite{Yurke:1987} suggested that canonical quantisation be completed with Dirac's procedure \cite{Dirac:1950,Dirac:1959}. Notably, Werner and Drummond \cite{Werner:1991} derived the first canonical Hamiltonian in a mix of flux variables for discrete elements, and charge fields for transmission lines. 

However, these efforts did not have continuity, and the flux-field Lagrangian (\ref{eq:TL_Lag_Phi}) became the standard tool for  canonical quantisation of superconducting circuits \cite{Devoret:1997,Paladino:2003,Blais:2004,Houck:2008,Bourassa:2009,Clerk:2010,Koch:2010,Filipp:2011,Bourassa:2012,Bergenfeldt:2012,Peropadre:2013,Nigg:2012,Solgun:2014,Sundaresan:2015,Solgun:2015,Malekakhlagh:2016,Mortensen:2016,Roy:2016,Vool:2017,Malekakhlagh:2017,Roy:2018,ParraRodriguez:2018}. Some issues arose in the precise derivation of the Hamiltonian \cite{Blais:2004,Houck:2008,Filipp:2011,Sundaresan:2015,Bourassa:2012,Peropadre:2013,Roy:2016,Vool:2017,Roy:2018}, in particular regarding the presence or otherwise \cite{Paladino:2003,Bourassa:2009,Bergenfeldt:2012,Bamba:2014,Mortensen:2016,Gely:2017,Malekakhlagh:2017} of divergences in Lamb-shifts~\cite{Lamb:1947}, radiative decays~\cite{Weisskopf:1930,Malekakhlagh:2017}, and effective couplings~\cite{Filipp:2011}. Correct solutions to these points have also been presented  \cite{Paladino:2003,Bourassa:2009,Bergenfeldt:2012,Gely:2017,Malekakhlagh:2017,Nigg:2012,Solgun:2014,Solgun:2015}. These divergences were actually artificial and avoidable artifacts, and the role of the flux-field Lagrangian as a predictive tool was restored \cite{ParraRodriguez:2018}. However, to date the flux-field Lagrangian approach has proven unable to handle the quantisation of {general} nonreciprocal systems. For a detailed example see below in subsection~\ref{sec:adhoc-description}.

\subsection{Reduced space operators}
Let us now review the standard separation of variables analysis (i.e.,  the normal mode construction for TLs) for the successful quantisation of  Lagrangian (\ref{eq:TL_Lag_Phi}) in a form  relevant for our extension. The analysis of (\ref{eq:TL_Lag_Q}) is completely analogous. In essence, one looks for an expansion of the fields in terms of an orthonormal basis for the relevant Hilbert space, with the coefficients of the expansion becoming the new dynamical variables to be quantised. Such a basis is determined by  a self-adjoint differential operator ${\cal L}$, acting on a domain in the Hilbert space of multicomponent functions of position with as many components as lines, namely $N$. For clarity, we will denote the multicomponent functions of position as $\bU(x)$ while we keep the notation $\bPhi(x,t)$ for the dynamical fields. The boundary conditions of the problem at hand are part of the definition of the domain, such as, as we assume in this section, open-boundary conditions  ($\mDelta\bPhi'(0,t)\equiv\mDelta\bPhi_0'=0$) at one end of the lines. In the case of interest, the action of the  operator is
\begin{equation}
	\mL \bU=-\mDelta \bU''\,.
      \end{equation}

This operator is positive, and we denote its (generalised) eigenvalues as $\omega^2$, choosing $\omega$ to be nonnegative real. As the equation is multicomponent, degeneracies might arise, depending on the symmetry of the circuit, and we use a discrete degeneracy index $\lambda$ to denote degenerate eigenstates, ${\cal L}\bU_{\omega\lambda}=\omega^2\bU_{\omega\lambda}$, orthonormalised with respect to the natural inner product. Since the differential expression and the boundary conditions of the operator ${\cal L}$ are real, the operator is real, and a real basis can always be chosen.

 We expand the flux fields in the eigenbasis,  $\bPhi=\int_{\mR_+}d\Omega\,F_{\omel}\bU_{\omel}$ and substitute in  the Lagrangian  to obtain $L_{\bPhi}=\frac{1}{2} \int d\Omega \left(\dot{F}_\omel^2-\omega^2F_\omel^2\right)$
where $\Omega=(\omega,\lambda)$, and  $\int d\Omega\equiv\sum_{\lambda}\int_{\mR_+} d\omega$. Now the set $F_\omel(t)$ are the new dynamical variables. There is no obstacle to carrying out the Legendre transformation, resulting in the Hamiltonian $H_{\bPhi}=\frac{1}{2}\int d\Omega \left(\Pi_\omel^2+\omega^2F_\omel^2\right)$, identical for both the flux-field and charge presentations,  where the canonical momenta are $\Pi=\partial L_{\bPhi}/\partial \dot{F}$. The process of canonical quantisation is now straightforward, by promoting the conjugate pairs of variables to quantum operators, and using annihilation and creation forms one recovers the standard expression
\begin{eqnarray}
	\hat{H}_{\bPhi}=\sum_{\lambda=1}^{N} \int d\omega\,\hbar \omega\,\hat{a}_\omel^\dag \hat{a}_\omel\equiv \hat{H}_{\bQ},\label{eq:TL_LU_Ham}
\end{eqnarray}
where we have discarded the infinite constant term  $\int d\Omega\frac{1}{2}$. The degeneracy index sum has been made explicit here. With the boundary conditions for one end of a set of half-lines as above, all frequencies are equally degenerate, all with the maximal degeneracy that equals the number of lines $N$.

\subsection{Double space operator}
Let us now apply the same procedure to the  telegrapher's Lagrangian (\ref{eq:TL_Lag_ds}) directly, without first eliminating the nondynamical components. Crucially, this involves a \emph{doubled}
space, with $\bW(x)$ the new multicomponent functions, now with $2N$ components. For later convenience we arrange $\bW$ as a doublet of two $N$ component functions, $\bU$ and $\bV$, corresponding, respectively, to fluxes and charges. We know, however, that only a maximum of  $N$ different combinations of these components can be physical. Consider now the operator defined by ${\cal L}\bW=-\mDelta\bW''$ on the elements of its domain. The compact notation signifies that $\mDelta$ applies to both $\bU$ and $\bV$ components, i.e., we use the symbol $\mDelta$ also for the block diagonal matrix $\mathrm{diag}(\mDelta,\mDelta)$. The boundary conditions, for simplicity, will again be open boundary conditions at one end of the lines, $\mDelta\bU'_0=\bV_0=0$.

  This new operator is again positive and real, and we shall use the corresponding real eigenbasis $\bW_\ome$, (generalised) orthonormal with respect to the inner product
\begin{eqnarray}
	\langle\bW_1,\bW_2\rangle=\int_{\mcl{I}} dx\, \left[\bU_1^T  \bU_2+\bV^T_1\mDelta^{-1}\bV_2\right]\,.\label{eq:Inner_product_ds}
\end{eqnarray}
The degeneracy index, which we will denote by $\epsilon,\epsilon'$, now runs up to $2N$. This is the degeneracy in this case of $N$ \emph{independent} and semiinfinite transmission lines. Other configurations, with different boundary conditions for finite TLs and different propagation velocities for each line, will generally present with smaller degeneracies.

\subsection{Telegrapher's symmetry}
We  now consider the duality of Eq. \eqref{eq:dualityblock}. This is in fact a consequence of  the electromagnetic duality symmetry \cite{Silberstein:1907,Calkin:1965,Jackson:1999}  given by the exchange of electric and magnetic fields in a TEM mode \cite{Pozar:2009}. As is well known, electromagnetic duality is not local when expressed in terms of the fundamental potential vector field, and we should not expect that here it be a mere rotation of flux and charge fields. In fact Eqs. \eqref{eq:dualityblock} and \eqref{eq:solduality} show explicitly that it is not local.

To make use of this duality of descriptions,  we introduce the {\it telegrapher's} operator $\mT$, acting on the elements of the domain of ${\cal L}$ as
\begin{equation}
	\mT\bW=-i\begin{pmatrix}
	\bV'\\\mDelta\bU'
	\end{pmatrix},\label{eq:TL_T_op_def}
      \end{equation}
where $\bW^T=(\bU^T,\bV^T)$. This is a discrete symmetry that commutes with the fundamental operator ${\cal L}$.  Furthermore,  $\mT^2={\cal L}$, and its spectrum is given by $\pm i\omega$. However, it is purely imaginary and, therefore, its action on the real basis $\bW_\ome$ cannot be merely a sign. Alternatively, notice that it interchanges non trivially flux and charge components. Thus, on each degeneracy eigenspace it will act as $\mT\bW_\ome=\omega \mt_{\epsilon\epsilon'}\bW_{\omega\epsilon'}$, with $\mt_{\epsilon\epsilon'}\delta_{\omega\omega'}=\langle\bW_{\omega'\epsilon'},\mT\bW_{\omega\epsilon}\rangle$. The matrix $\mt$  must be therefore imaginary and idempotent. In fact, by rearranging the real basis in each degeneracy subspace, it can always be written as $\mt=-\sigma_y\otimes\mone_N$, with $\mone_N$ the identity matrix on $N$ dimensional vectors, see Appendix~\ref{App_sec:Self_adjoint_op} for further details of self-adjoint operators in the doubled space with more general boundary conditions.

\subsection{Lagrangian and Hamiltonian in double space}
\label{sec:lagr-hamilt-double}
We now have all the necessary tools to complete the analysis in the doubled case. Expand the flux and charge doublet in the orthonormal basis $\bW_\ome$,
  \begin{equation}
    \label{eq:Phi_Q_expansion}
    \begin{pmatrix}
      \bPhi\\ \bQ 
    \end{pmatrix}=\int d\Omega\,X_\ome\bW_\ome=\int d\Omega\,X_{\ome}\begin{pmatrix}
    \bU\\\bV
  \end{pmatrix}_{\ome}\,,
\end{equation}
where $\int d\Omega$ stands for sum over both frequencies and degeneracy indices, and $X_\ome(t)$ are the time dependent mode variables.
 The Lagrangian (\ref{eq:TL_Lag_ds}) becomes
\begin{eqnarray}
	L=\frac{1}{2}\int d\Omega\, \dot{X}_{\ome}(\dot{X}_{\ome}-i\omega\mt_{\epsilon'\epsilon}X_{\omega\epsilon'}).\label{eq:Lag_TL_modes_ds}
\end{eqnarray}
Notice that the matrix $\mt$ that implements the telegrapher's symmetry appears explicitly in the Lagrangian. It now behoves us to prove that, albeit different at a first glance from $L_{\bPhi}$ in Eq. \eqref{eq:TL_Lag_Phi} above, equation (\ref{eq:Lag_TL_modes_ds}) produces equivalent dynamics for the flux and charge variables.  For instance, the last term is apparently breaking TRS, and indeed that would be the case in general. However, we have indicated from the outset that for	reciprocal circuits TRS does hold. We prove in Appendix~\ref{sec:time-reversal} that (\ref{eq:Lag_TL_modes_ds}) does preserve TRS in reciprocal contexts.

On writing $\mt$ as $-\sigma_y\otimes\mathbb{1}_N$, it is convenient to rearrange the degeneracy indices accordingly, with $u/v$ being the indices for the action of the Pauli matrix, while the $N$ dimensional factor corresponds to an index $\lambda,\lambda'$. In other words, the single degeneracy index $\epsilon$, ranging from 1 to $2N$, is replaced by a doublet $\alpha\lambda$, with $\alpha$ taking values $u$ and $v$, and $\lambda$ from 1 to $N$. Let us denote $X_{\omega(u\lambda)}$ by $F_{\omega\lambda}$ and $X_{\omega(v\lambda)}$ by $G_\omel$.
The Lagrangian is rewritten as 
\begin{eqnarray}
L=\frac{1}{2}\int d\Omega\,\left[ \dot{F}_{\omel}^2+\dot{G}_{\omel}^2+\omega\left(\dot{G}_\omel  F_{\omel}-\dot{F}_\omel G_\omel\right)\right],\nonumber
\end{eqnarray}
where now the implicit sum runs over $\int d\Omega\equiv\sum_{\lambda=1}^{N}\int d\omega$.
Notice two salient facts: the kinetic term is nondegenerate, and this Lagrangian is amenable to Legendre transform to provide us with a Hamiltonian, first, and, second, the term in parenthesis is of magnetic nature. Its antisymmetry is directly inherited from the antisymmetry of $\mt$, hence from the telegrapher's duality symmetry. In fact, the Hamiltonian reads
\begin{eqnarray}
	H=\frac{1}{2}\int d\Omega\,\left[\left(\Pi_\omel+\frac{\omega}{2}G_{\omel}\right)^2+\left(P_\omel-\frac{\omega}{2}F_{\omel}\right)^2\right],\nonumber
\end{eqnarray}
with  two pairs of conjugated variables with Poisson brackets $\{F_{\omel},\Pi_{\omega'\lambda'},\}=\{G_{\omel},P_{\omega'\lambda'},\}=\delta_{\omega\omega'}\delta_{\lambda\lambda'}$. The identification of the structure as being magnetic reveals to us the dynamical content of this Hamiltonian. Namely, under the canonical transformation $\tilde{F}_\omel=\frac{1}{2}F_\omel-\frac{1}{\omega}P_\omel$, $\tilde{\Pi}_\omel=\Pi_\omel+\frac{\omega}{2}G_\omel$, $\tilde{G}_\omel=\frac{1}{2}G_\omel-\frac{1}{\omega}\Pi_\omel$ and $\tilde{P}_\omel=P_\omel+\frac{\omega}{2}F_\omel$, 
the Hamiltonian reaches its final form
\begin{eqnarray}
	&H&=\frac{1}{2}\int d\Omega\,\left(\tilde{\Pi}_\omel^2+\omega\tilde{F}_{\omel}^2\right)\nonumber\\
	&&\equiv_{\mathrm{q.}}\sum_{\lambda}^{N}\int d\omega\, \hbar\omega\, \hat{a}_{\omel}^\dag \hat{a}_\omel\,.\label{eq:H_quantised_ds_modes}
\end{eqnarray}
Although at first sight in (\ref{eq:Lag_TL_modes_ds}) there are $2N$ modes in each degeneracy space, our treatment reveals that half of those are \emph{nondynamical} ($\tilde{G}_\omel$ and $\tilde{P}_\omel$), in that they have no evolution under the physical Hamiltonian (\ref{eq:H_quantised_ds_modes}). As we pointed out above, the telegrapher's Lagrangian is redundant, in that it presents a gauge symmetry that has no dynamics, and we have made this apparent by means of the canonical transformation. Once this result has been achieved and the nondynamical variables in phase space have been set to zero, thus making a concrete gauge choice, quantisation follows in the same manner as before. 

In summary, to this point we have shown that the telegrapher's Lagrangian (\ref{eq:TL_Lag_ds}) is amenable to quantisation, and that it controls the correct number of degrees of freedom. The most standard way to quantise circuits, nicely summarised in Devoret's \cite{Devoret:1997,Vool:2017}, begins by writing a set of equations of motion (EOMs), and continues by finding a Lagrangian to which these E-L equations correspond. Notice that if we were to start from the telegrapher's equations we would first need to reduce them to the wave equation in the standard (reduced) approach and then understand those as the E-L equations of the Lagrangians presented as $L_{\bPhi}$ and $L_{\bQ}$ in (\ref{eq:TL_Lag_Phi}) and (\ref{eq:TL_Lag_Q}). We are following a similar  route: we start directly from the telegrapher's equations and understand them as E-L for $L_{\mathrm{TG}}$. The redundancy is not made to disappear in  the EOMs (configuration space) but rather in the Hamiltonian (phase space). This we have achieved by introducing a systematic analysis, based on the telegrapher's duality operator. Although it is not necessary to use the doubled space for the description of the systems above,  we shall  use this systematic procedure to handle the quantisation of a system for which it will prove crucial, with the inclusion of ideal nonreciprocal elements.

\section{Connection through ideal nonreciprocal elements}
\begin{figure}[h!]
	\centering
	\includegraphics[width=0.5\linewidth]{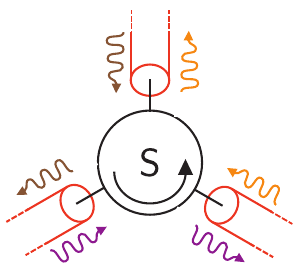}
	\caption{Transmission lines coupled to an ideal circulator. We assume a common ground plane for the lines and circulator ports.}
	\label{fig:S_3ports}
\end{figure}

Let us briefly recap the fundamentals of linear lossless nonreciprocal devices \cite{Duinker:1959,Tellegen:1948,Carlin:1964}, and explain their inclusion in Lagrangian models with transmission lines, e.g., a 3-port circulator connected directly to three TLs as in Fig. (\ref{fig:S_3ports}).

A linear ideal NR system can be generically described by a unitary, non-symmetric (frequency-independent) scattering matrix $\mS$. This matrix relates the amplitude and phase of input and output signals $\bsb{b}^{\mathrm{out}}=\msf{S}\bsb{b}^{\mathrm{in}}$, or voltages and currents at the ports
\begin{equation}
	(1-\msf{S})\dot{\bPhi}_0=R(1+\msf{S})\dot{\bQ}_0,\label{eq:NR_constitutive}
\end{equation}
 where $R>0$ is a reference resistance, and $\mS$ is a constant matrix. We assume here and in the rest of the article that all frequency-dependence of $\mS(\omega)$ could have been extracted in a network of capacitors and inductors  \cite{Carlin:1964,Newcomb:1966}. Such more involved networks require more careful treatment of the double-space variables \cite{ParraRodriguez:2021,ParraRodriguezPhD:2021}, and we restrict here the analysis to ideal NR devices. When neither $+1$ nor $-1$ are eigenvalues of $\mS$ \cite{Carlin:1964,ParraRodriguez:2019} the constitutive Eq. (\ref{eq:NR_constitutive}) can be simplified to admittance $\dot{\bQ}_0=\bar{\mY}\dot{\bPhi}_0$ and impedance $\bar{\mZ}\dot{\bQ}_0=\dot{\bPhi}_0$ equations, where $\bar{\mY}$ and $\bar{\mZ}$ are skew-symmetric real matrices. We consider here nonreciprocal systems with an immittance description, see Appendix~\ref{AppSec:Degenerate_cases} for the discussion of degenerate cases.
 
\subsection{Obstacles in the reduced description}
As we shall now show, there are fundamental obstacles in treating  systems of transmission lines coupled with ideal nonreciprocal elements using just flux (or charge) variables. One kind of problem is common both to networks of lumped elements and to systems of transmission lines, when in presence of nonreciprocity, and we first address it. Consider the simplest ideal nonreciprocal element, the gyrator \cite{Tellegen:1948}. It is a system that does not store energy, and in this regard it resembles a transformer. The difference is that an ideal transformer expresses a constraint between flux variables (or charge variables), while the constitutive equations of a gyrator are a constraint between flux on one port and charge on the other. It is also the case for a general nonreciprocal element that it provides us with constraints between fluxes and charges. Transformer constraints can be used to reduce the number of variables in a Lagrangian expressed in terms of flux variables, but this is not feasible for  nonreciprocal elements.

There is a second family of issues that pertains to transmission lines coupled to lumped elements, and the literature dealing with it in the reciprocal case has been extensive. A first possiblity for the description of these situations is to construct a Hamiltonian and hence canonical quantisation directly in terms of fields. This approach does work for transmission lines with open or closed boundary conditions, for example, by computing conjugate momentum densities. The classical field theory analysis provides us with a canonical symplectic form also in the simple case, and field quantisation goes through. This fails however (among others) in the case of capacitive coupling with flux variables, since the momentum density will of necessity be a distribution, with a Dirac delta term. We would then have to face the problem that the product of distributions is generically ill defined, and would require regularisation~\cite{Bamba:2014,Malekakhlagh:2016}. Therefore one must either discretise the system, as is done in much of the literature, and then take carefully a continuum limit, or use a mode expansion. For details on the momentum density and canonical quantisation of fields in this context,  see~\cite{ParraRodriguez:2018}.

Passing therefore to the analysis of the traditional mode expansion for flux fields coupled capacitively, notice that the conservation of charge (current) at the coupling leads to $\mathsf{\Delta}\boldsymbol{\Phi}_0'= \mA\left(\ddot{\boldsymbol{\Phi}}_0-\ddot{\boldsymbol{\Phi}}_L\right)$, where $\mA=\mathsf{c}_\delta^{-1/2}\mathsf{C}\mathsf{c}_\delta^{-1/2}$ is a (rescaled) capacitance coupling matrix and $\boldsymbol{\Phi}_L$ denotes other, discrete, degrees of freedom. Clearly, unless there is perfect cancellation of $\ddot{\boldsymbol{\Phi}}_0$ and  $\ddot{\boldsymbol{\Phi}}_L$ at all times, no normal mode analysis is feasible with standard Sturm--Liouville boundary conditions for the wave equation part, as made explicit in Sec. 2.1 of  \cite{ParraRodriguez:2018}. Therefore an extension of the usual Sturm--Liouville method is required. We have shown elsewhere that it exists and have provided explicit constructions, involving just flux variables. They rely crucially on the rescaled capacitance matrix being symmetric, since it is required for the construction of an inner product \cite{ParraRodriguez:2018}.

Let us now consider in this regard transmission lines coupled at one end through an ideal (frequency independent) nonreciprocal element with an admittance representation, when described with flux variables. Explicitly, we assess the possibility of direct canonical quantisation in terms of flux fields, first, and of the possible need to define an extension of the Sturm Liouville analysis, second. The Lagrangian is
\begin{equation}
	\label{eq:simpleYflux}
	L=L_{\boldsymbol{\Phi}}+\frac{1}{2}\dot{\boldsymbol{\Phi}}_0^T\mathsf{Y}\boldsymbol{\Phi}_0\,,
\end{equation}
with $\mathsf{Y}$ the skew-symmetric  rescaled admittance matrix, $\mathsf{Y}=\mathsf{c}_\delta^{-1/2}\bar{\mathsf{Y}}\mathsf{c}_\delta^{-1/2}$. Time-reversal symmetry is explicitly broken by the last term, clearly, and this will pose problems on its own, as we shall presently see. Setting that aside for the moment, a direct attempt at building the Hamiltonian density and the Hamiltonian in terms of the flux field and a conjugate momentum density runs into the same type of issue regarding regularisation~\cite{Bamba:2014,Malekakhlagh:2016} of products of distributions as the previous case.  Explicitly, the conjugate momentum densities for Lagrangian \eqref{eq:simpleYflux} are
\begin{equation}
\boldsymbol{Q}(x,t)=\dot{\boldsymbol{\Phi}}(x,t)+\frac{1}{2}\delta(x)\mathsf{Y}\boldsymbol{\Phi}(0,t),\,\nonumber
\end{equation}
and a Legendre transformation in these terms does not exist. It is again the case that either  a regularisation recipe is necessary, and to date none has  been presented in the literature, or, in the alternative, a mode expansion is required. We will follow this second way.

Before we do so, let us examine again in this case the issue of nondynamical components. The Euler--Lagrange equations derived from this Lagrangian are given by the wave equations (\ref{eq:wave_Phi}) and the current conservation at the end
\begin{eqnarray}
	\mathsf{\Delta}\boldsymbol{\Phi}'_0&=& \mathsf{Y}\dot{\boldsymbol{\Phi}}_0\,.\nonumber
\end{eqnarray}

This is a well posed problem, and explicit solutions can be found for particular configurations. For instance, a system with two lines, identity velocity matrix, and a gyrator $\mathsf{Y}=i\sigma^y$ rescaled admittance matrix has the general solution
\begin{equation}
	\label{eq:example}
	\boldsymbol{\Phi}\to
	\begin{pmatrix}
		f(t-x)+g(t+x)\\ g(t-x)-f(t+x)
	\end{pmatrix}
	\,.
\end{equation}
By inspection we notice that it is dynamically equivalent to a wave equation in just one of the lines, thus making apparent again the redundancy in the description of the system.

We now turn back to the identification of adequate normal modes and the corresponding expansion. It immediately becomes clear that no Sturm--Liouville boundary conditions are available to us in this flux representation. The reason is that  the charge conservation equation $\mathsf{\Delta}\boldsymbol{\Phi}'_0= \mathsf{Y}\dot{\boldsymbol{\Phi}}_0$ involves the time derivative, and would lead, under separation of variables, to a frequency dependent boundary condition for the normal form, $\mathsf{\Delta}\boldsymbol{U}_0'=-i\omega \mathsf{Y}\boldsymbol{U}_0$. The technique used in \cite{ParraRodriguez:2018} to address a similar issue in capacitive coupling, namely the consideration of a radically different kind of self-adjoint operator, is now not available to us, because of the skew-symmetry of the admittance, as opposed to the symmetry of the capacitance matrix $\mA$ presented above. 

That the extension of Sturm--Liouville theory presented in \cite{ParraRodriguez:2018} for capacitive coupling is not available can also be understood from an analysis of the relevant mathematical literature and from the breaking of time-reversal invariance in nonreciprocal systems. In the  system under description, Eq. \eqref{eq:simpleYflux}, time-reversal symmetry is broken, and this breaking is localised on a boundary. This implies that the charge conservation law at the boundary will, under separation of variables, present a term linear in frequency. On the other hand, the eigenvalues of the relevant second order differential operator, were it to exist, depend on frequency as $\omega^2$. The mathematical literature for eigenvalue problems in which the boundary condition involves the eigenvalue, following the seminal work of Walter \cite{Walter:1973} and Fulton  \cite{Fulton:1977}, provides us with self-adjoint operators if the boundary condition depends \emph{linearly} on the eigenvalue. If time-reversal symmetry is broken on the boundary, however, the boundary condition depends on the \emph{square root} of the eigenvalue. Therefore the construction we relied on for capacitive coupling (in the flux description) is no longer available. 

This set of arguments, both those general to networks and to transmission lines and those specific to couplings of transmission lines to lumped elements, demonstrates that a description only in terms of fluxes is incapable to provide us with the systematic canonical quantisation we desire.

This does not mean, at all, that canonical quantisation cannot be achieved. For instance, direct inspection of Eq. \eqref{eq:example} shows that the second component can be understood as conjugate to the first one. Indeed, in that simple situation a Sturm--Liouville separation of variables is accessible by describing one line in fluxes and the second line in charges, and the procedure provides us with the desired result. This comes about because charge conservation reads in this case, in terms of components,
\begin{eqnarray}
  \delta_1\Phi_1'(0,t)&=& \mathsf{Y}_{12}Q'_2(0,t)\,,\nonumber\\
   \mathsf{Y}_{21}\dot{\Phi}_1(0,t)&=&\dot{Q}_2(0,t)\,.\nonumber
\end{eqnarray}

Redefining space-time units so as to have $\delta_1=1$ and $\mathsf{Y}_{12}=-\mathsf{Y}_{21}=1$, in correspondence to example above, Eq. \eqref{eq:example}, we  define a  differential operator $\mathcal{L}$ acting on the doublet $\bW=
\begin{pmatrix}
  U&V
\end{pmatrix}^T$ as $\mL \bW=-\bW''$, with boundary conditions $U(0)=-V(0)$ and $U'(0)=V'(0)$. It is self-adjoint when using the standard inner product, and expansion on the corresponding basis leads directly to diagonalised quantisation. Observe that this analysis crucially depends on using a configuration space with some flux and some charge variables. In fact, of the two lines one is  described in terms of flux and the other one in terms of charge. See App. \ref{AppSec:TLs_gyrator} for the full exposition of this example.

In generalizing this idea to a larger number of transmission lines, the assignment of flux or charge character to individual transmission lines is not straightforward, in fact in general not even possible. In a nutshell, this flux or charge assignment would require identifying the change of variables that transforms the  skew-symmetric matrix $\mY$ into its canonical form, given by $2\times2$ independent skew-symmetric blocks, and a block of zeroes. Once the change of variables has been computed, the effective pairs associated with the canonical form should be given a flux and charge character, one each per pair. The issue arises in that the description of the original lines is now given by a superposition of these fluxes and charges, and  if  couplings with  capacitive and inductive aspects are also present there is no way of clearly describing them with the new combination of variables. Namely, the energy corresponding to capacity would be neither clearly kinetic nor potential, and similarly for inductors. If they are linear it might be possible to disentangle these issues for some cases. If there are nonlinearities, however, canonical quantisation is fraught with problems.

This obstacle to generalizing the alternating flux/charge assignment above provides us with the idea for a solution, that we now prove is indeed adequate and systematic: describe \emph{all} lines with \emph{both} flux and charge variables, and provide a process to identify the dynamical degrees of freedom, now in phase space, thus eliminating the inherent redundancy.

\subsection{Solution in the doubled space}
\label{sec:solut-doubl-space}
Having looked into the barriers associated with the description in terms only of fluxes, we present a general solution based on the doubled-space description. Let us consider an $N$-port nonreciprocal element connected to $N$ semi-infinite lines as in Fig. (\ref{fig:S_3ports}). The Lagrangian of the system in the doubled space can be written as 
\begin{align}
L &= L_{\mathrm{TG}}+\frac{1}{2}\left\{
\begin{aligned}
&\dot{\bQ}_0^T\bPhi_0+ \dot{\bPhi}_0^T\msf{Y}\bPhi_0,\\
&\dot{\bPhi}_0^T\bQ_0+ \dot{\bQ}_0^T\msf{Z}\bQ_0,
\end{aligned}
\right. \label{eq:TL_YZ_Lag_ds}
\end{align}
working with rescaled flux and charge fields, and matrices $\mY$ for admittance and $\mZ=\sqrt{\mcd}\bar{\mZ}\sqrt{\mcd}$ for impedance. The constitutive equation (\ref{eq:NR_constitutive}) written in admittance form is the Euler-Lagrange boundary condition equation in the corresponding Lagrangian (\ref{eq:TL_YZ_Lag_ds}). We concentrate here on the admittance case, as the impedance one is analogous. The relevant operator ${\cal L}$ acts on a doublet $\bW$ of $N$ component functions, $\bU$ and $\bV$, belonging to a domain restricted by the conditions
\begin{equation}
  \label{eq:Domain_L_Y}
  \bV_0 = \mY \bU_0,\qquad \mathrm{and}\qquad\mDelta\bU'_0=\mY \bV'_0\,,
\end{equation}
and, as before, 
\begin{equation}
	\mL\bW=-\begin{pmatrix}
		\mDelta\bU''\\\mDelta\bV''
	\end{pmatrix}.
\end{equation}
 The corresponding inner product is again (\ref{eq:Inner_product_ds}), and the telegrapher's duality symmetry, defined as before, but now on the current domain, mantains its crucial properties, all the way to its representation by means of $\mt$ and its properties. See Appendix~\ref{App_sec:Self_adjoint_op} for further details on the operator and its eigenbasis.

Since the boundary  terms cancel on expanding the Lagrangian \eqref{eq:TL_YZ_Lag_ds} in an eigenbasis $\bW_\ome$ of ${\cal L}$, \begin{eqnarray}
&&	\dot{\bQ}_0^T\bPhi_0+ \dot{\bPhi}_0^T\msf{Y}\bPhi_0=\nonumber\\
	&&=	\int d\Omega d\Omega' \dot{X}_{\ome}\bU_{\ome}^T(0)(\mY^T+\mY)\bU_{\omep}(0)X_{\omep}=0,\nonumber
\end{eqnarray}
due to $\mY$ being skew-symmetric, the same steps as before lead us to a properly quantised Hamiltonian, of the form of Eq. \eqref{eq:H_quantised_ds_modes}, and,  as before, there are  $N$ degrees of freedom per frequency mode. Again, even if we have introduced an apparent redundancy in the description of the system, there is a clear and systematic identification of the dynamical variables, and the number of these is the correct one. Observe that even though the $N$ lines are constrained by the nonreciprocal element, the degeneracy is still that of the case of $N$ independent lines.

We have throughout stressed that time-reversal symmetry (TRS)
holds in the reciprocal case, and we have also pointed out that
the nonreciprocity introduced by the boundary terms of Eq. (\ref{eq:TL_YZ_Lag_ds}) explicitly breaks TRS. On the other hand, the final form of the quantised Hamiltonian we present is identical for both the reciprocal case (in both the doubled and the reduced presentations) and the nonreciprocal case, namely $\sum_{\lambda=1}^N\int
d\omega\,\hat{a}^\dag_{\omega\lambda}{a}^\dag_{\omega\lambda}$. Surprising as this might look at first sight, there is no contradiction. In both cases the transformation of the flux and charge fields under time reversal is dictated by their physical meaning. As we explicitly show in Appendix~\ref{sec:time-reversal}, the sequence of changes of variables for the reciprocal case time reversal is diagonal in frequency, and realises as an antiunitary mapping $\hat{a}$ to $-\hat{a}^\dagger$ and $\hat{a}^\dagger$ to $-\hat{a}$, but this is no longer the case when the nonreciprocal terms are present, and modes are mixed. Therefore the diagonal quantum Hamiltonian is not invariant under time reversal, although it has the same form as the reciprocal one.

Notice that the quantisation of nonreciprocal systems is fraught with difficulties, as attested by the scarcity of actual constructions in the literature. Here we have  illustrated in particular  how our technique solves and clarifies the issue of time-reversal symmetry, for definiteness. Other issues, such as possible divergences in Lamb shifts and other measurable quantities, are also addressed with our technique, as we will portray in the next section.

In summary, we have presented a systematic quantisation procedure for ideal nonreciprocal devices coupled to transmission lines, by using the doubled presentation of the telegrapher's equation and discarding the uncoupled nondynamical sector.

\section{Nonlinear networks}
\begin{figure}[h]
	\centering
	\includegraphics[width=.95\linewidth]{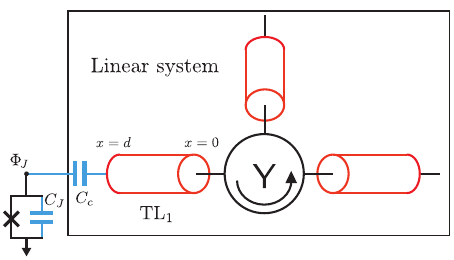}
	\caption{Josephson junction connected to transmission-line resonators and a $\mY$-circulator. The linear sector of the Hamiltonian is described by an infinite set of harmonic oscillators.}
	\label{fig:FL_NL_network}
\end{figure}

In order to show the power of our approach, we now expand the construction presented in \cite{ParraRodriguez:2018} to describe also networks of ideal circulators connected to finite-length transmission lines with capacitive connections to Josephson junctions as in Fig. \ref{fig:FL_NL_network}. Particular as this example might be, it still captures all the essential difficulties of much more general situations. More concretely, similar constructions to what we are now going to present would apply to capacitive/inductive insertions in the line or in connection to more general meshes, see \cite{ParraRodriguez:2018} for a catalogue. 

In addition to the boundary conditions (\ref{eq:Domain_L_Y}) at the circulator, the analysis of the circuit in Fig. \ref{fig:FL_NL_network} requires imposing current conservation at the $x=d$ endpoint of the first transmission line, $\mathrm{TL}_1$. This extra condition is not of Sturm--Liouville type, however, and, as shown in \cite{ParraRodriguez:2018}, it is necessary to consider an operator that does not act merely on the multicomponent Hilbert space of the lines, but rather on the direct sum of that Hilbert space and a boundary finite-dimensional Hilbert space, such that its elements are of the form $\mW=\left(\bW,w\right)$. In the case at hand $w$ is a real number. In this expanded space, with the corresponding inner product, ${\cal L}$ acts on its domain, defined by the boundary conditions (\ref{eq:Domain_L_Y}) and $(\mDelta\bU'_d)_\perp=(\bV_d)_\perp=0$
together with the restriction $w=\alpha(\bsb{n}\cdot \bU_d)$,
as  $\mL\mW=(-\mDelta\bW'', -(\bsb{n}\cdot\mDelta\bU'_d))$, where $\bsb{n}$ is an $N$ dimensional vector and $\perp$ denotes orthogonal to $\bsb{n}$. $\alpha$ is a free parameter which can conveniently be set to the value $C_s/c=C_c C_J/[c(C_c + C_J)]$, such that the Hamiltonian does not have TL mode-mode couplings ($A^2$ diamagnetic term), see \cite{ParraRodriguez:2018,Malekakhlagh:2017} and Appendix~\ref{AppSec:Example} for further details. For simplicity, we have assumed open boundary conditions in the endpoints of the other lines ($x=d$).

We now follow the same procedure as before, namely (1)  expansion in the eigenbasis of this operator to obtain a Lagrangian in new dynamical variables, for which the Legendre transformation is well defined, and (2)  identification of the dynamical sector of proper dimension (equivalent to the  elimination of nondynamical variables in the previous section).  We arrive  at the quantum Hamiltonian 
\begin{eqnarray}
	\hat{H}&=&\frac{\hat{Q}_J^2}{2C_J}-E_J\cos(\hat{\varphi}_J)+\sum_{n,\lambda}^{\infty,N} \hbar\omega_n\, \hat{a}_{n\lambda}^\dag \hat{a}_{n\lambda} \nonumber\\
	&&+\xi \hat{Q}_J\sum_{n,\lambda}\left[r_{n\lambda} \hat{a}_{n\lambda}+r^*_{n\lambda} \hat{a}_{n\lambda}^\dagger\right],\nonumber
\end{eqnarray}
where $r_{n\lambda}=\sqrt{\frac{\hbar\omega_n}{2}}  (u_{n u\lambda}+i u_{n v\lambda})$, with $u_{n\epsilon}=\bsb{n}\cdot(\bU_{n\epsilon})_d$  the eigenfunctions at the coupling point $\hat{\Phi}_1(x, t)=\sum_{n,\epsilon}\hat{X}_{n\epsilon}(t)(\bsb{n}\cdot\bU_{n\epsilon}(x))$, and $\xi=\frac{C_c}{(C_c+C_J)\sqrt{c_\delta}}$. As suggested above,  the computation of Lamb-shifts for this Hamiltonian, which will be proportional to $\chi\propto \sum_{n,\lambda}|r_n|^2/\omega_n=\hbar/(2\alpha)<\infty$, proves to be convergent, since, as previously shown in \cite{Malekakhlagh:2017,Gely:2017,ParraRodriguez:2018} for reciprocal networks, $u_{n\epsilon}\propto 1/n$ when $\omega_n\rightarrow\infty$. 

To our knowledge, this is the first exact Hamiltonian description combining transmission lines, nonreciprocal elements and Josephson junctions. In contrast to the procedures described in \cite{ParraRodriguez:2019} for pure lumped-element networks with nonreciprocal elements, the methods here take advantage of the inherent properties of the transmission lines, and allow us to construct a unique eigenvalue problem (for a self-adjoint operator) to not just quantise but further diagonalise the linear sector even in the presence of the nonreciprocal elements. Furthermore, one may safely take the infinite-length limit of the lines (continuum spectrum) and still find meaningful (nonrelativistic) Hamiltonians predicting divergence-free Lamb shifts or effective couplings. 

\section{Conclusions}
We have put forward a consistent quantisation procedure for  superconducting circuits modeled in a redundant flux-charge description with a corresponding double-space basis. The apparent redundancy is eliminated by making use of a duality symmetry. Such a doubled basis becomes mandatory for the systematic identification and quantisation of the dynamical  degrees of freedom in circuits with ideal nonreciprocal devices, e.g. circulators connected to transmission lines. We have applied the theory to a circuit with a Josephson junction connected to a circulator through a transmission line, deriving a Hamiltonian in which there is no mode-mode coupling among the dressed normal modes. This class of quantum Hamiltonians are free of artificial divergence issues, incorporating their inherent Lorentz-Drude kind of cutoff in a predictive manner. Further forthcoming work is required to expand this description to  more general nonreciprocal devices with frequency dependence, where the problem reduces to correctly analysing TLs coupled through the full triplet of capacitors, inductors and ideal NR elements \cite{ParraRodriguezPhD:2021,ParraRodriguez:2021}.

\begin{acknowledgments}
We acknowledge support from the Basque Government through grant IT986-16. A. P.-R. thanks Basque Government Ph.D. grant PRE-2016-1-0284, and enlightening discussions with C. Bruzzone, I. Arrazola, D. Huerga, B. Reina and L. Garc\'ia-\'Alvarez.
\end{acknowledgments}

\appendix

\section{Reduced space operator}
We formally define the differential operators for the half-line in the reduced space  by their domain
\begin{eqnarray}
\mD(\mL)&=&\left\{\bsb{E},\,\bsb{E} \in \mathbbm{C}^N\otimes AC^1(\mcl{I}),\right.\nonumber\\ &&\,\,\,\,\left.\msf{D}_{\bE}\bsb{E}_0+\msF_{\bE} \bsb{E}'_0=0 \right\},\label{eq:TL_LU_VE_def}
\end{eqnarray}
with $\bE$ either $\bU$ or $\bV$, $\msf{D}_{\bU}=\msF_{\bV}=0$, $\msF_{\bU}=\mDelta$, and $\msD_{\bV}=\mone$, and by  their action on elements of the domain $\bE\in\mD(\mL)$
\begin{equation}
\mL \bE=-\mDelta \bE''.
\end{equation}
We associate the notation $\bU$ with fluxes and $\bV$ with charges throughout. We use for convenience two different inner products, $\langle \bU_1,\bU_2\rangle=\int_{\mcl{I}}dx\, \bU_{1}(x)\bU_2(x)$ for the flux and $\langle \bV_1,\bV_2\rangle=\int_{\mcl{I}}dx\, \bV_{1}(x)\mDelta^{-1}\bV_2(x)$ for the charge presentation, with which one can construct orthonormal bases for the different spaces.  We denote with $\bE_{\omega\lambda}$ the generalised eigenvectors of the operator $\mL$, where $\omega\in\mR_+$ is a continuous parameter and $\lambda$ is a discrete degeneracy index with a maximal range given by the number of lines $N$, that solve the Sturm-Liouville eigenvalue problems $\mL \bE_{\omel}=\omega^2 \bE_{\omel}$, with positive eigenvalue $\omega^2$. The reality of the coefficients of the differential expressions and the boundary conditions translates into the reality of the operator, $\mL=\mL^*$, and thus a real basis can always be chosen.

\section{Double space operator for nonreciprocal elements}
\label{App_sec:Self_adjoint_op}

The domain of the self-adjoint operator proposed for the mode decomposition of the admittance-described NR element connected to $N$ semi-infinite lines is
\begin{eqnarray}\label{eq:NonRecBCs}
\mcl{D}(\mL)&=&\left\{\bW(x)\in AC^1(\mcl{I})\otimes\mathbbm{C}^{2N},\right.\nonumber\\
            &&\left.\bV_0 = \mY \bU_0,\,\mDelta\bU'_0=\mY \bV'_0\right\}\,.\end{eqnarray}
To describe finite length lines we would need to supplement with adequate boundary conditions the definition of the domain. In what follows we systematically understand $\bW$ to be a doublet of $N$ component functions,
\begin{equation*}  \bW=
\begin{pmatrix}  \bU\\ \bV  \end{pmatrix}
\,. \end{equation*}
The operator acts on elements of its domain  as 
\begin{eqnarray}
\mL \bW &=&-\mDelta\bW''=\begin{pmatrix}  -\mDelta\bU''\\ -\mDelta\bV''  \end{pmatrix}\,.\label{eq:L_op_appendix}
\end{eqnarray}
It can be easily checked that under the inner product 
\begin{eqnarray}
\langle\bW_1,\bW_2\rangle&=&\int_{\mcl{I}}dx \,\bW_1^\dag \mUpsilon \bW_2\label{eq:inner_product_lines_only}
\end{eqnarray}
with $\mUpsilon=\text{diag}(\mone,\mDelta^{-1})$, the operator is self-adjoint, as follows. For $\bW_2$ to be in $\mD(\mL^\dag)$, there needs to exist a $\bW_3$ such that \begin{eqnarray}
&\langle \mL\bW_1,\bW_2\rangle-\langle\bW_1,\bW_3\rangle=0
\end{eqnarray}
holds $\forall\bW_1\in\mD(\mL)$, in which case one defines $\bW_3=\mL^\dag\bW_2$. Now, by integration by parts  
\begin{align}
&\langle \mL\bW_1,\bW_2\rangle-\langle\bW_1,-\mDelta\bW_2''\rangle=\nonumber\\
&\bV_1'(0)^\dag\left[\mY^T\bU_2(0)+\bV_2(0)\right]\nonumber\\
&-\bU_1^\dag(0)\left[\mDelta \bU_2'(0)+\mY^T\bV_2'(0)\right]\,.\nonumber
\end{align}
Naturally, for finite length lines additional terms related to the other boundary will appear, as we shall show explicitly in an example in Appendix \ref{AppSec:TLs_gyrator}. Here we concentrate on the ideal lumped nonreciprocal element. Since $\bV_1'(0)$ and $\bU_1(0)$ are not determined by the boundary conditions, only by their relations to other quantities, in order for the RHS to be zero for all $\bW\in{\cal D}({\cal L})$ the terms in square brackets have to be zero. We thus see that the domains of ${\cal L}$ and ${\cal L}^\dag$ match. The requirement of absolutely continuous derivative is necessary for the applicability of integration by parts.
\subsection{Completeness}
The operator defined above (in this case  in the interval $\mcl{I}=\mR^+$) is a monotone (accretive) operator, i.e.,  for all  $\bW\in \mcl{D}(\mL)$ we have 
\begin{eqnarray}
\langle \bW, \mL\bW\rangle&=&\int_{\mR^+} dx\,(\bW^\dag)'\mUpsilon\mDelta\bW'> 0,\nonumber
\end{eqnarray}
 as $\mUpsilon\mDelta$ is a real positive symmetric matrix. Furthermore, by standard arguments of Sturm--Liouville operators, it is readily shown to be maximal monotone and thus its domain is dense in $L^2(\mcl{I})\otimes\mathbbm{C}^{2N}$. It follows that its eigenbasis is a basis for the whole Hilbert space. We shall use this fact to expand general configurations in this basis.
\subsection{$\mT$ Symmetry}
Let us define a transformation $\mT$ on the domain of $\mL$ by 
\begin{eqnarray}
\mT\bW&=&-i\begin{pmatrix}
\bV'\\ \mDelta\bU'
\end{pmatrix}\nonumber\,.
\end{eqnarray}
Since $W\in\mD(\mL)$ is absolutely continuous, this is well defined. Now, the crucial property of this transformation is that applied to eigenvectors that obey $\mL\bW=\omega^2\bW$ it is the case that $\mT\bW$ belongs to the domain of $\mL$ and is again an eigenvector of $\mL$ with the same eigenvalue. Thus, extended by linearity, the two of them commute \[\mL\left[\mT\bW\right]=\mT\left[\mL\bW\right]\,.\]
It follows that, unless $\mT$ acts trivially on the $\omega^2$ eigenspace, that eigenvalue is degenerate. Thus, generically, the presence of this symmetry give rise to an at least two-fold degeneracy of the eigenvalues. This transformation implements electromagnetic duality in the present context.

\subsection{Orthonormal eigenbasis}
Let us  now compute with specificity  an orthonormal basis for the spectral decomposition of the differential operator in the case of semi-infinite TL. The general solution for the system of ordinary differential equations $\mL\bW=\omega^2\bW$ is 
\begin{eqnarray}
\bW_{\ome}(x)&=&\mcl{N}\left[\cos\left(\omega x\mDelta^{-\frac{1}{2}}\right)\begin{pmatrix}
\mDelta^{-\frac{1}{2}}\be_c\\ \mDelta^{\frac{1}{2}}\br_c
\end{pmatrix}\right.\nonumber\\
&&\left.+ \sin\left(\omega x\mDelta^{-\frac{1}{2}}\right)\begin{pmatrix}
\be_s\\ \br_s
\end{pmatrix}\right].\notag
\end{eqnarray}
We have introduced an additional normalisation parameter $\mcl{N}$ for later convenience.
Introducing the general solution in the  boundary conditions  one can fix  $2N$ constants of the general solution as 
\begin{eqnarray}
\br_c&=&\tilde{\mY}\be_c,\notag\\
\be_s&=&\tilde{\mY}\br_s,\notag
\end{eqnarray}
where  we have rescaled the admittance matrices with the velocity matrix, i.e.,  $\tilde{\mY}=\mDelta^{-\frac{1}{2}}\mY\mDelta^{-\frac{1}{2}}$. In general it might be the case that   $\br_c$, $\be_r$, and the free $\br_s\equiv \br$ and $\be_c\equiv \be$  depend on frequency. In the example at hand the spectrum is continuous, and to determine the (generalised) eigenbasis we now demand (generalised) orthonormality, 
\begin{equation}
\langle\bW_{\ome},\bW_{\omep}\rangle=\delta_{\omega\omega'}\delta_{\epsilon\epsilon'}.
\end{equation}
Inserting the general solution with the restrictions set by the boundary conditions, and making use of the identities
\begin{eqnarray}
\int_{\mR^+}dx\,\cos(\omega x\mDelta^{-\frac{1}{2}})\cos(\omega' x\mDelta^{-\frac{1}{2}})&=&\frac{\pi}{2}\delta_{\omega\omega'}\mDelta^{\frac{1}{2}},\notag\\
\int_{\mR^+}dx\,\sin(\omega x\mDelta^{-\frac{1}{2}})\sin(\omega' x\mDelta^{-\frac{1}{2}})&=&\frac{\pi}{2}\delta_{\omega\omega'}\mDelta^{\frac{1}{2}},\notag\\
\int_{\mR^+}dx\,\cos(\omega x\mDelta^{-\frac{1}{2}})\sin(\omega' x\mDelta^{-\frac{1}{2}})&=&\frac{1}{2}\mDelta^{\frac{1}{2}}\times\dots\nonumber\\
\dots\times\mcl{P}\left(\frac{1}{\omega+\omega'}-\frac{1}{\omega-\omega'}\right).
\end{eqnarray}
we rewrite the orthonormality conditions as
\begin{equation}
\label{eq:morthono}
\frac{\pi}{2}\mcl{N}_\ome^2\begin{pmatrix}
\be\\\br
\end{pmatrix}_\ome^T \mcM\begin{pmatrix}
\be\\\br
\end{pmatrix}_{\omega\epsilon'}=\delta_{\epsilon\epsilon'}\,,
\end{equation}
where 
\begin{eqnarray}
\mcM&=&\begin{pmatrix}
\mM&0\\0&\mM
\end{pmatrix}=\mcM^T,\nonumber\\
\mM&=& \mDelta^{-\frac{1}{2}}+\tilde{\mY}^T\mDelta^{\frac{1}{2}}\tilde{\mY},\nonumber
\end{eqnarray}
whence we see that $\mM$ is symmetric. In this way we have rewritten the orthonormality conditions as an algebraic problem, namely that of finding vectors that are orthogonal with respect to the quadratic form determined by $\mcM$. This is tantamount to its diagonalisation problem, clearly. Because of its structure, we can assert that  its   eigenvalues $m_\lambda$, are at least doubly degenerate. They are independent of $\omega$, in contrast to more general possibilities. Furthermore we do know that it is diagonalizable, which allows us to state that the degeneracy index $\epsilon$ runs from 1 to $2N$ for all eigenvalues $\omega$ of $\mL$. Each eigenvalue $m_\lambda$ of $\mcM$ is  associated with a pair of orthogonal eigenvectors $(\be^T, 0)_{\lambda}^T$ and $(0, \be^T)_{\lambda}^T$, also independent of $\omega$. These two sets allow us additionally  to separate the degeneracy index $\epsilon$ running from $1\leq\epsilon\leq 2N$ into a $\{u,\,v\}$ index and a $1\leq\lambda\leq N$ such that we write the orthonormal basis 
\begin{eqnarray}\label{eq:uvbasis}
&&\bW_{\omega (u\lambda)}=\sqrt{\frac{2}{\pi m_\lambda}}\cos\left(\omega x\mDelta^{-\frac{1}{2}}\right)\begin{pmatrix}
\mDelta^{-\frac{1}{2}}\be\\ \mDelta^{\frac{1}{2}}\tilde{\mY}\be
\end{pmatrix}_{\lambda},\nonumber\\
&&\bW_{\omega (v\lambda)}=\sqrt{\frac{2}{\pi m_\lambda}} \sin\left(\omega x\mDelta^{-\frac{1}{2}}\right)\begin{pmatrix}
\tilde{\mY}\be\\ \be
\end{pmatrix}_{\lambda}.
\end{eqnarray}

\subsubsection*{Telegrapher's matrix representation}
It follows from the details of the computation above that in this basis the matrix representation of the $\mT$ operator  has the explicit block shape
\begin{align}
\langle\bW_{\ome},\mT\bW_{\omep}\rangle&=\delta_{\omega\omega'}\omega\mt_{\epsilon'\epsilon}=\delta_{\omega\omega'}\omega\begin{pmatrix}
0&\mt_{uv}\\\mt_{vu}&0
\end{pmatrix}^T\nonumber\\
&=\delta_{\omega\omega'}\omega(\sigma_y\otimes\mone_{N}).\notag
\end{align}

\section{Time-reversal symmetry}
\label{sec:time-reversal}
Maxwell's equations are invariant under time reversal (TR). This property must be reflected in the effective equations for transversal-electromagnetic modes in transmission lines, i.e.,  in the telegrapher's equations. Since the charge fields are integrals of current intensities, that are odd with respect to time reversal, the charge fields will be even under  time reversal with respect to the origin of integration. On the other hand, the flux field will be odd, since the voltage, its derivative, is even. Thus, the time-reversal operation pertinent here is
\begin{equation}
	\label{eq:TRS1}
	\boldsymbol{\Phi}\to-\boldsymbol{\Phi}\,,\qquad  \boldsymbol{Q}\to\boldsymbol{Q}\,,\qquad t\to - t\,,
\end{equation}
and both the telegrapher's equations~(\ref{eq:telegraphers}) and  Lagrangian~(\ref{eq:TL_Lag_ds}) are invariant under this transformation.

Even more, the reduction to the wave equations and to the flux ($L_{\boldsymbol{\Phi}}$) and charge ($L_{\boldsymbol{Q}}$) Lagrangians preserves the symmetry. It is frequently the case that time reversal for these is presented simply in terms of $t\to-t$, since $L_{\boldsymbol{\Phi}}$ presents an additional $\mathbbm{Z}_2$ symmetry $\boldsymbol{\Phi}\to\boldsymbol{\Phi}$, but the actual, physical, TR transformation that one must investigate is \eqref{eq:TRS1}.

In the process of canonical quantisation we are here separating variables. The spatial part, which is determined by the self-adjoint operators we put forward, has no bearing on time-reversal symmetry or lack thereof, and it must be studied in the mode amplitudes. Let us thus first analyse TRS for the traditional separation of variables for $L_{\boldsymbol{\Phi}}$. The normal forms $\boldsymbol{U}_{\omega\lambda}$ are invariant under time reversal. Thus, the expansion in the eigenbasis $\bPhi(x,t)=\int_{\mR_+}d\Omega\,F_{\omel}(t)\bU_{\omel}(x)$ provides us with the TRS operation
\begin{equation}
	\label{eq:TRSforU}
	F_{\omel}\to-F_{\omel}\,,\qquad t\to-t\,.
\end{equation}
In the process of Legendre transforming we then have the full TRS operation now on phase space.
\begin{equation}
	\label{eq:TRSforFpi}
	F_{\omel}\to-F_{\omel}\,,\qquad \Pi_{\omel}\to\Pi_{\omel}\,,\qquad t\to-t\,.
\end{equation}
Therefore, the classical complex amplitudes $a_{\omel}$ are transformed as
\begin{equation}
	\label{eq:TRSforeasya}
	a_{\omel}\to - a_{\omel}^\dag\,,\qquad a_{\omel}^\dag\to - a_{\omel}\,,\qquad t\to-t\,.
\end{equation}

The realisation of the TRS operation \eqref{eq:TRS1} in the ``double space'' case of subsection II.B is a bit more involved. In fact, prima facie the Lagrangian (12) seems to break time-reversal symmetry, due to the presence of the $-i\omega\dot{X}\mathsf{t}X$ term. Let us analyse the TRS operation in the context of mode expansion, namely
\begin{align}
	\label{eq:TRSmodes}
	&\begin{pmatrix}
		\boldsymbol{\Phi}(t)\\ \boldsymbol{Q}(t)
	\end{pmatrix}\to  \begin{pmatrix}
	-\boldsymbol{\Phi}(-t)\\ \boldsymbol{Q}(-t)
\end{pmatrix}=\\
	&\qquad\qquad\qquad=\int\mathrm{d}\Omega\,X_{\omega\epsilon}(-t)
	\begin{pmatrix}
		-1&0\\0&1
	\end{pmatrix}
	\begin{pmatrix}
		\boldsymbol{U}\\
		\boldsymbol{V}
	\end{pmatrix}_{\omega\epsilon}\,,\nonumber
\end{align}
omitting the spatial dependencies. This is the general transformation for all the expansions and situations that we consider. We see then that the TRS operation transfers to the action of $-\sigma^z$ on the mode doublet $\boldsymbol{W}_{\omega\epsilon}$.

Crucially, $\sigma^z$ commutes with the operator $\mL$ in the reciprocal case. Notice that this involves both the commutation of the velocity matrix $\mathsf{\Delta}$ with $\sigma^z$ \emph{and} the fact that the boundary conditions required for the definition of the domain of $\mL$ are invariant under the action of $\sigma^z$. We can therefore diagonalise simultaneously $\mL$ and $\sigma^z$, and indeed that is exactly the form in which we have proceeded. Namely, in rewriting the degeneracy index $\epsilon$ as  $\alpha\lambda$ with $\alpha$ being either $u$ or $v$, we are choosing
\begin{equation}
	\label{eq:sigmazplus}
	\sigma^z\boldsymbol{W}_{\omega u\lambda}=\boldsymbol{W}_{\omega u\lambda}\quad\mathrm{and}\quad
	\sigma^z\boldsymbol{W}_{\omega v\lambda}=-\boldsymbol{W}_{\omega v\lambda}\,.
\end{equation}
It follows that in this case we can write the TRS operation on the mode amplitudes as
\begin{eqnarray}
	\label{eq:xtransf}
	F_{\omega\lambda}(t)=X_{\omega u\lambda}(t)&\longrightarrow& -F_{\omega\lambda}(-t)\,,\\
	G_{\omega\lambda}(t)=X_{\omega v\lambda}(t)&\longrightarrow& G_{\omega\lambda}(-t)\,.\nonumber
\end{eqnarray}
From here it flows that their conjugate momenta transform as
\begin{eqnarray}
	\label{eq:momtrans}
	\Pi_{\omega\lambda}(t)&\longrightarrow& \Pi_{\omega\lambda}(-t)\,,\\
	P_{\omega\lambda}(t)&\longrightarrow& -P_{\omega\lambda}(-t)\,,\nonumber
\end{eqnarray}
and the symplectic transformation we use to eliminate unphysical degrees of freedom respects this structure, namely $\tilde{F}$ is odd, $\tilde{\Pi}$ is even, etc. Therefore the final Hamiltonian is invariant under the TRS induced on the final variables, as expected.

Let us now carry out the corresponding analysis in a more abstract form, also applicable to the nonreciprocal case. As stated above, time reversal is the operation indicated in Eqn. \eqref{eq:TRSmodes} in all cases, both reciprocal and nonreciprocal. The crucial issue at this juncture is that $\sigma^z$ does \emph{not} commute with the nonreciprocal operator $\mathcal{L}$ defined in Appendix B, because the domain determined by the boundary conditions, Eqn. \eqref{eq:NonRecBCs}, is not invariant under the action of $\sigma^z$. Thus, the following matrix is not diagonal:
\begin{eqnarray}
	\label{eq:zdef}
	\mSigma_{\omega\epsilon,\omega'\epsilon'}&=&\left\langle\boldsymbol{W}_{\omega\epsilon},\sigma^z \boldsymbol{W}_{\omega'\epsilon'}\right\rangle\,.
\end{eqnarray}
When the degeneracy index $\epsilon$ is written as above in the form $(u/v)$$\lambda$, we introduce and compute the matrix 
\begin{eqnarray}\label{eq:SigmaMatrix}
	\mSigma_{\omega\lambda,\omega'\lambda'}
	&=&
	\delta_{\omega\omega'}\delta_{\lambda\lambda'}\sigma^z\label{eq:zdef_uv}\\ &&
	+\frac{2}{\sqrt{m_\lambda m_\lambda'}}
	\delta_{\omega\omega'}\boldsymbol{e}_\lambda^T
	\tilde{\mY}\mathsf{\Delta}^{1/2}\tilde{\mY}
	\boldsymbol{e}_{\lambda'}\,\mone\nonumber\\ &&-\frac{4}{\pi\sqrt{m_\lambda
			m_\lambda'}}
	\mcl{P}\frac{1}{\omega^2-\left(\omega'\right)^2} \begin{pmatrix}
		0&\omega'\\ \omega &0\end{pmatrix}
	\boldsymbol{e}_\lambda^T\tilde{\mY}\boldsymbol{e}_{\lambda'}\,.\nonumber
\end{eqnarray}
From (\ref{eq:zdef}) we write explicitly the transformation rule for the mode amplitudes $X_{\omega\epsilon}$, namely
\begin{equation}
	\label{eq:xTRS}
	X_{\omega\epsilon}(t)\to -\int d\Omega'\, \mSigma_{\omega\epsilon,\omega'\epsilon'} X_{\omega'\epsilon'}(-t)\,.
\end{equation}
Once we have reached equation (\ref{eq:Lag_TL_modes_ds}), we can analyse its transformation under time reversal. We will be using a compact notation, in which Eq. \eqref{eq:xTRS} is written as
  \begin{equation}
	\label{eq:compactxTRS}
	X(t)\longrightarrow X^\tau(t)= - \mSigma X(-t)\,.
\end{equation}
The Lagrangian (\ref{eq:Lag_TL_modes_ds}) transforms as 
\begin{equation}
	\label{eq:1ltrans}
	\dot{X}^T\left(\dot{X}-i\omega\mathsf{t}X\right) \to \left(\dot{X}^\tau\right)^T\mSigma^T\left(\mSigma\dot{X}^\tau+i\omega\mathsf{t}\mSigma X^\tau\right)\,.
\end{equation}
Using the properties of the basis, and the fact that $\sigma^z$ is self adjoint with respect to the inner product,  $\mSigma\mSigma^T=\mone$. In the reciprocal case ($\tilde{\mY}=0$), $	\mSigma_{\omega\lambda,\omega'\lambda'}$ from (\ref{eq:zdef_uv}) is diagonal in frequency and in the $\lambda$ degeneracy index, and equals $\sigma^z$ in the $u/v$ index. Since $\mathsf{t}$ is, in essence, $\sigma^y$, and $\sigma^z\sigma^y\sigma^z=-\sigma^y$, we again prove TRS in this formalism for the reciprocal case.

Let us now examine the TR transformation in phase space. A general TR transformation stems from a rule in configuration space, as above, which induces an anticanonical transformation in phase space. Restricting ourselves to linear TR transformations, as is the case here, the canonical momenta transform with minus the inverse of the transpose. Explicitly, with the notation introduced in Eq. \eqref{eq:compactxTRS} but not making the argument explicit,
  we have
  \begin{align}
    \label{eq:TRPhaseSpace}
    X^\tau_\Omega&=-\Sigma_{\Omega\Omega'}X_{\Omega'}\,,\nonumber\\
    \left(\Pi^\tau_X\right)_\Omega&=  \Sigma^{-1}_{\Omega'\Omega}\left(\Pi_X\right)_{\Omega'}\,.
  \end{align}

Thus time reversal mixes the modes we are analyzing in the nonreciprocal case. In other words,  this transformation structure, Eqns. \eqref{eq:TRPhaseSpace} and \eqref{eq:zdef} , that is determined by the physical TR rule \eqref{eq:TRS1},  will generally mix the modes under study when the system is nonreciprocal.

\section{TLs connected through a gyrator: adhoc vs systematic quantisation}
\label{AppSec:TLs_gyrator}
\begin{figure}[h]
	\centering
	\includegraphics[width=1\linewidth]{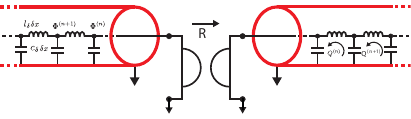}
	\caption{Two transmission lines coupled through an ideal gyrator.}
	\label{fig:TL_Gyr_TL}
\end{figure}

We present here two possible ways to quantise canonically  two transmission lines connected by an ideal gyrator, see Fig. \ref{fig:TL_Gyr_TL}. First, in the manner mentioned in the main text, we use an adhoc description of the system in terms of a charge field in one transmission line and a flux field in the second. We then use the alternative and systematic procedure described in this article that allows us to quantise circuits with transmission lines connected through ideal nonreciprocal elements with arbitrary number of ports, and which permits a more flexible and convenient description of couplings of the transmission lines to further lumped nonlinear elements (like Josephson junctions).

Without loss of generality and for the sake of simplicity we work in the homogeneous case (velocities and characteristic impedances of both TLs are equal, whence $\mDelta=v^2\mone_2$, and with open and short boundary conditions for the flux-described and charge-described TL, respectively.  It is important to note that the (dynamical sector) degeneracy of each frequency will not be  $N=2$, as would seem to follow from the analysis of section \ref{sec:solut-doubl-space}, but just 1. This follows from the presence of the boundary conditions: effectively, here we have one non-free TL of length $2L$ with boundary conditions, thus reducing the degeneracy.
\subsection{Adhoc description}\label{sec:adhoc-description}
The equations of motion for the circuit in Fig. \ref{fig:TL_Gyr_TL} can be readily written as
\begin{align}
	\ddot{\Phi}(x,t)&=v_p^2 \Phi''(x,t),\,\,\,
	\ddot{Q}(x,t)=v_p^2 Q''(x,t),\,\label{eq:FTL_gyrator_QTL_waveq}\\
	\dot{\Phi}(0,t)&=-\frac{R}{v_p Z_0}	\dot{Q}(0,t),\label{eq:FTL_gyrator_QTL_bc2}\\  
	\Phi'(0,t)&=\frac{Z_0}{v_p R}Q'(0,t),\label{eq:FTL_gyrator_QTL_bc1}\\
	\Phi'(L,t)&=Q'(L,t)=0,\label{eq:FTL_gyrator_QTL_bc3}
\end{align}
where the characteristic impedance and propagation velocity of both lines are $Z_{0}=\sqrt{l/c}$ and $v_p=1/\sqrt{lc}$, respectively. These equations of motion follow from the Lagrangian
\begin{align}
	L =&\frac{1}{2} \int_{\mcl{I}} dx\,\left[\dot{\Phi}(x,t)^2 -v_p^2(\Phi'(x,t))^2\right]\nonumber\\
	&+ \frac{1}{2}\int_{\mcl{I}} dx\,\left[\frac{1}{v_p^2}\dot{Q}(x,t)^2 -(Q'(x,t))^2\right]\nonumber\\
	&+\frac{1}{R}\phi_3\dot{\Phi}(0,t)/\sqrt{c}+\phi_3 \dot{Q}(0,t)/v_p\sqrt{l} \label{eq:FTL_gyrator_QTL_Lag},
\end{align}
where the last line contains a Lagrange multiplier $\phi_3$ to enforce the appropiate nonreciprocal constraints on the boundaries. The finite interval is $\mcl{I}=(0,L)$. Under separation of variables
\begin{equation}
	\Gamma(x,t)=\begin{pmatrix}
		\Phi\\
		Q
	\end{pmatrix}(x,t)=X(t)\begin{pmatrix}
		U(x)\\
		V(x)
	\end{pmatrix},
\end{equation}
the equations (\ref{eq:FTL_gyrator_QTL_waveq}-\ref{eq:FTL_gyrator_QTL_bc3}) give us
\begin{align}
	\frac{\ddot{X}}{X}(t)&=v_p^2 \frac{U''}{U}(x)=v_p^2 \frac{V''}{V}(x)=-\omega^2,\nonumber\\
	U(0)&=-\frac{R}{v_p Z_0}V(0),\quad U'(0)=\frac{Z_0}{v_p R}V'(0),\nonumber\\
	U'(L)&=V'(L)=0.\nonumber
\end{align}
We are going to use the eigenbasis of the self-adjoint differential operator $\mL$ to expand functions of position. This operator $\mL$  acts on doublets $\bW=(U(x), V(x))^T \in AC^1(\mcl{I})\otimes\mathbb{C}^2$ as $\mL\bW=-\mDelta\bW''$. Since we desire expansions in an eigenbasis, and further that the boundary conditions of the physical system are properly expressed,  we ensure  selfadjointness with a domain  $\mD(\mL)$ determined by the boundary conditions of the physical system
\begin{align}
	\mD(\mL)=\{&\bW\in AC^1(\mcl{I})\otimes\mathbb{C}^2,\nonumber\\
	& \msf{A} \bW(0)+ \msf{B} \bW'(0)=0,	\nonumber\\
	&\msf{C} \bW(L) + \msf{D} \bW'(L)=0\},\nonumber
\end{align}
with the matrices being
\begin{eqnarray}
	\msf{A}&=&\begin{pmatrix}
		1 & R/v_p Z_0\\ 0 & 0
	\end{pmatrix}, \,\, \msf{B}=\begin{pmatrix}
		0 & 0\\ 1 &- Z_{0}/v_p R
	\end{pmatrix},\nonumber\\
	\msf{C}&=&\begin{pmatrix}
		0 & 1\\ 0 & 0
	\end{pmatrix}, \,\, \msf{D}=\begin{pmatrix}
		0 & 0 \\1 & 0
	\end{pmatrix}.\nonumber
\end{eqnarray}
The normal mode wavenumbers are the solutions of the transcendental equation 
\begin{align}
k\cos(k L) \sin(k L) =0,
\end{align}
with solutions $k_{n}=\frac{n \pi}{L}$, $k_{n+\frac{1}{2}}=\frac{(2n+1)\pi}{2L}$, and $n\in \mathbb{N}$, including zero (because the eigenvalues are actually $\omega^2=v_p^2k^2$). Bearing in mind that the inner product is
\begin{equation*}
  \langle \bW_a,\bW_b\rangle=\int_\mcl{I} dx\, \left[U_a(x)U_b+\frac{1}{v_p^2}V_a(x)V_b(x)\right]\,,
\end{equation*}
the correspondingly normalised associated  eigenvectors are
\begin{align}
	\bW_{n}(x)&=\mcl{N} \cos\left(k_{n} x\right)\begin{pmatrix}
		R\\ -v_p Z_0
	\end{pmatrix}\nonumber,\nonumber\\
	\bW_{n+\frac{1}{2}}(x)&=\mcl{N} \sin\left(k_{n+\frac{1}{2}} x\right)\begin{pmatrix}
	Z_0\\v_p R
\end{pmatrix},\nonumber\\
\mcl{N}&=\sqrt{\frac{2}{L(R^2+Z_0^2)}}.\nonumber
\end{align}
Introducing the decomposition above described into the Lagrangian (\ref{eq:FTL_gyrator_QTL_Lag}) with the standard inner product we derive 

\begin{align}
	L=\,&\frac{1}{2}\sum_{a,a'}\dot{X}_a \dot{X}_{a'} \langle \bW_a,\bW_{a'}\rangle\nonumber\\
	&-\frac{v_p^2}{2}\sum_{a,a'}X_a X_{a'} \langle \bW_a',\bW_{a'}'\rangle\nonumber\\
	=\,&\frac{1}{2}\sum_{a}\left(\dot{X}_a^2-\omega_a^2 X_a^2\right),\nonumber
\end{align}
where the indices $a,a'$ here run over both sets of indices, integer and halfinteger. The conjugate variables are derived from the Legendre transformation $\Pi_a=\partial L/\partial \dot{X}_a = \dot{X}_a$ and the final Hamiltonian is
\begin{align}
	H&=\sum_a \frac{\Pi_a^2}{2}+\frac{\omega_a^2 X_a^2}{2}\equiv_{\mathrm{q.}}\sum_a \omega_a a_a^\dagger a_a,\nonumber\\
	&=\sum_n \omega_{n}a_n^\dag a_n + \omega_{n+\frac{1}{2}} a_{n+\frac{1}{2}}^\dag a_{n+\frac{1}{2}}\nonumber
\end{align}
where we have finished the analysis by quantizing canonically the conjugate dynamical variables, and writing $H$ in terms of creation and annhilation operators. We have discarded the zero point contribution. The eigenenergies are not degenerate.

Let us now study the behaviour of this system under time reversal. The $\Gamma$ doublet transforms as $-\sigma^z\Gamma$. We have to compute, as before, the transformation matrix
\begin{equation*}
  \mathsf{\Sigma}_{ab}=\left\langle \bW_a,\sigma^z\bW_b\right\rangle\,.
\end{equation*}
Explicitly,
\begin{align}
  \label{eq:sigmaadhoc}
  \mathsf{\Sigma}_{n\,m}&= \delta_{nm}\frac{R^2-Z_0^2}{R^2+Z_0^2}\,,\nonumber\\
  \mathsf{\Sigma}_{n\,m+1/2}&=\mathsf{\Sigma}_{m+1/2\,n}\nonumber\\
  &= \frac{4Z_0R}{\pi\left(R^2+Z_0^2\right)}\frac{m+1/2}{(m+1/2)^2-n^2}\,,\nonumber\\
  \mathsf{\Sigma}_{n+1/2\,m+1/2}&= \delta_{nm}\frac{Z_0^2-R^2}{R^2+Z_0^2}\,.
\end{align}
Using some standard summation formulae it is readily  checked that this matrix is orthogonal, in fact idempotent. Under time reversal, therefore,
\begin{align}
  \label{eq:TRadhocPhase}
  X^\tau_a&= -\mathsf{\Sigma}_{ab}X_b\,,\\
  \Pi^\tau_a&=\mathsf{\Sigma}_{ab} \Pi_b\,.\nonumber
\end{align}
The Hamiltonian is thus expressed in these new variables as
\begin{equation*}
  H= \frac{1}{2}\sum_a\left(\Pi^\tau_a\right)^2+\frac{1}{2}\sum_{bc}\left(\sum_a \omega^2_a\mathsf{\Sigma}_{ba}\mathsf{\Sigma}_{ca}\right)X_c^\tau X_b^\tau\,.
\end{equation*}
Observe thus that under time reversal there is mode-mode coupling, as promised.
\subsection{Doubled description}
We apply now the double-space solution for this particular case. In  App. \ref{App_sec:Self_adjoint_op} we have described  the generic properties of this construction, exemplified with $N$ semi-infinite TLs. Here we shall particularise to $N=2$, and include boundary conditions at a finite length $L$. For ease of comparison we use the same parameters as in subsection \ref{sec:adhoc-description}.  The domain of the self-adjoint operator is now
\begin{align}\label{eq:domain_L_TL_Gyr_TL_ds}
	\mcl{D}(\mL)=&\left\{\bW(x)\in AC^1(\mcl{I})\otimes\mathbbm{C}^{4},\right.\nonumber\\
	&\left.\bV_0 = \mY \bU_0,\,\mDelta\bU'_0=\mY \bV'_0,\right.\nonumber\\
	&U_1'(L)=V_1(L)=U_2(L)=V_2'(L)=0\left.\right\}\,,
\end{align}
where the rescaled admittance matrix is 
\begin{align}
	\mY=c^{-1/2}\bar{\mY}c^{-1/2}=v_p\frac{Z_0}{R}\begin{pmatrix}
		0&1\\-1&0
	\end{pmatrix}
\end{align}
and the velocity matrix is $\mDelta=v_p^2\mone$.
Note that the boundary conditions at the endpoint $x=L$ are those of open- /short- circuit conditions for the left/right TL. The operator $\mL$ acts on quadruplets ($\mathbb{C}^4$) of functions as $\mL\bW=-v_p^2\bW''$. The flux/charge pairs in each line will be expanded as
\begin{equation*}
  \begin{pmatrix}
    \Phi_1\\
    \Phi_2\\
    Q_1\\
    Q_2
  \end{pmatrix} =\sum_a X_a(t) \bW_a(x)\,,
\end{equation*}
with $\bW_a(x)$ the eigenfunctions of $\mL$. To compare with the previous subsection, consider the pair $\{\Phi_1,Q_2\}$,

The eigenfrequencies are obtained from the transcendental equation
\begin{align}
	k^2 \cos^2(k L) \sin^2(k L)=0,
\end{align}
with $\omega= v_p k$, and are again determined by  $k_{n}=\frac{n \pi}{L}$ and  $k_{n+\frac{1}{2}}=\frac{(2n+1)\pi}{2L}$. As demonstrated in the text for the generic situation, the eigenvalues $\omega_n^2=v_p^2k_n^2$ and $\omega_{n+\frac{1}{2}}^2=v_p^2k_{n+\frac{1}{2}}^2$ are all degenerate. In fact, in this case they all present degeneracy two. The corresponding eigenvectors can be organised as
\begin{align}
	\bW_{n,u}(x)&=\mcl{N}\cos(k_{n} x)\begin{pmatrix}
		R\\0\\0\\-v_p Z_0
	\end{pmatrix}\nonumber,\\
	\bW_{n+\frac{1}{2},u}(x)&=\mcl{N}\cos\left(k_{n+\frac{1}{2}} x\right)\begin{pmatrix}
	0\\R\\v_pZ_0\\0
	\end{pmatrix},\nonumber\\
	\bW_{n,v}(x)&=\mcl{N}\sin(k_n x)\begin{pmatrix}
		0\\-Z_0\\v_pR\\0
	\end{pmatrix},\nonumber\\
	\bW_{n+\frac{1}{2},v}(x)&=\mcl{N}\sin\left(k_{n+\frac{1}{2}} x\right)\begin{pmatrix}
	Z_0\\0\\0\\v_pR
\end{pmatrix},\nonumber\\
\mcl{N}=\frac{1}{R}&\sqrt{\frac{2}{L(1+v_p^{-2}\msf{Y}_{21}^2)}}=\sqrt{\frac{2}{L(R^2+Z_0^2)}}.\nonumber
\end{align}
We use the notation $u/v$ for the degeneracy index, without additional indices $\lambda$. The choice presented here matches the structure of Eq. \eqref{eq:uvbasis}, with $\be\propto
\begin{pmatrix}
  1&0
\end{pmatrix}^T$ for integer and  $\be\propto
\begin{pmatrix}
  0&1
\end{pmatrix}^T$ for half-integer indices.

The duality operator $\mT$, defined by Eq. \eqref{eq:TL_T_op_def}, acts a $\mT \bW_{a,u}=i\omega_{a}\bW_{a,v}$, and $\mT \bW_{a,v}=-i\omega_{a}\bW_{a,v}$ for $a$ either integer or halfinteger. Thus, it is represented in each degeneracy space (with the order $(uv)$) as
\begin{align}
	\msf{t}=-\sigma_y\,.\nonumber
\end{align}
In this case there is no identity tensorial factor, since degeneracy is just two for each energy.

We expand the flux and charge fields all together using this basis, 
\begin{align}
	\begin{pmatrix}
		\bPhi\\\bQ
	\end{pmatrix}(x,t)&=\sum_{a, \epsilon} X_{a\epsilon}(t)\begin{pmatrix}
	\bU\\\bV
\end{pmatrix}_{a\epsilon}(x),\nonumber
\end{align}
where $a$ denotes both integers and halfintegers ($\omega_{n}$ and $\omega_{n+\frac{1}{2}}$), while $\epsilon$ is either $u$ or $v$. Let us denote  $X_{au}=F_{a}$ and $X_{av}=G_{a}$ the two different coordinates per frequency $\omega_a$. The double-space Lagrangian (\ref{eq:Lag_TL_modes_ds}), with the particular boundary conditions fixed in this section, becomes when written in terms of modes in 
\begin{align}
	L=\frac{1}{2}\sum_{a}\,\left[ \dot{F}_{a}^2+\dot{G}_{a}^2+\omega_a\left(\dot{G}_{a}  F_{a}-\dot{F}_{a} G_{a}\right)\right].\nonumber
\end{align}
After the steps of  Legendre transformation, canonical transformation, elimination of nondynamical variables in phase space,  and canonical quantisation of the dynamical subspace, as  described in the main text in section \ref{sec:lagr-hamilt-double}, we arrive yet again at the quantised Hamiltonian up to the zero point energy,
\begin{align}
	H=\sum_{n}\omega_n a_n^\dag a_n+ \omega_{n+\frac{1}{2}} a_{n+\frac{1}{2}}^\dag a_{n+\frac{1}{2}}.
\end{align}
As expected, there is no degeneracy in this case in the dynamical space.

The two analyses performed in this appendix permit to draw the following conclusions: 
\begin{itemize}
\item The double-space method allows us to find systematically the convenient duality rotation (mixing of the charge and flux fields) for connections of TLs through NR ideal elements with an arbitrary number of ports. Such transformation is performed in the classical phase-space before the canonical quantisation program is applied. In the particular example of the circuit in Fig. \ref{fig:TL_Gyr_TL} in the previous subsection, we have exploited the special symmetry in the circuit and described just with one type of field each line to perform an adhoc canonical quantisation. That is no longer possible, for instance, when treating an ideal 3-port circulator. 
\item In addition, the double-space method has the flexibility to introduce couplings to nonlinear lumped degrees of freedom in the most convenient manner, typically as nonlinear potentials, e.g. capacitive coupling to a Josephson junction (see the analysis of the circuit in Fig. \ref{fig:FL_NL_network}) or inductive coupling to a phase-slip junction
\begin{align}
	L=\,&L_{\mathrm{TLs}}+ L_{\mY}(0)\nonumber\\
	&+\frac{C_c}{2}(\dot{\Phi}_i(L) - \dot{\Phi}_J)^2+\frac{L_c}{2}(\dot{Q}_j(L) - \dot{Q}_P)^2\nonumber\\
	&- E_J\cos(2\pi\Phi_J/\Phi_0)-E_{P}\cos(\pi Q_{P}(L,t)/e)\nonumber\\
	&+\dots\nonumber
\end{align}
\end{itemize}

\section{$\msf{S}$ matrix degenerate case}
\label{AppSec:Degenerate_cases}
Here, we briefly discuss how the analysis should be updated when  admittance or impedance matrices do not exist for the ideal nonreciprocal element. The constitutive equation for the nonreciprocal system is
\begin{equation}
(1-\mS)\dot{\bPhi}_0=R^{-1}(1+\mS)\dot{\bQ}_0.
\end{equation}
Let us consider the case where we want an admittance description as for the impedance case we have the dual problem. As discussed in \cite{ParraRodriguez:2019}, we may project on the space of eigenvalue $-1$ to find the constraint 
\begin{equation}
\mP_1\dot{\bPhi}_0=0,
\end{equation}
from where one can reduce the number of free coordinates to $N-1$. An admittance equation can be written as 
\begin{equation}
\tilde{\bQ}_0=\tilde{\mY}\tilde{\bPhi}_0,\notag
\end{equation}
where the matrix $\tilde{\mY}=R^{-1}(\mQ_1(1+\mS)\mQ_1)^{-1}\mQ_1(1-\mS)\mQ_1$, and the flux $\tilde{\bPhi}_0$ and charge $\tilde{\bQ}_0$ vectors have $N-1$ entries. The domain of the operator to treat this case is updated to be 
\begin{eqnarray}
\mcl{D}(\mL)&=&\left\{\bW(x)\in AC^1(\mcl{I})\otimes\mathbbm{C}^{2N},\right.\nonumber\\
&&\left.\,\mQ_1\bV_0 = \tilde{\mY}\mQ_1 \bU_0,\,\mQ_1\mDelta\bU'_0=\tilde{\mY} \mQ_1\bV'_0\right\}.\notag
\end{eqnarray}
\section{Analysis of the circuit in Fig.~\ref{fig:FL_NL_network}}\label{AppSec:Example}
Here we show the full computation of the Hamiltonian for the circuit in Fig. 3 of the main text. To do that, we need to enlarge the differential operator \cite{ParraRodriguez:2018} to describe networks of ideal circulators connected to transmission lines which may have capacitive connections to nonlinear networks. The domain of the new operator with an additional boundary for the first line is
\begin{align}
\mcl{D}(\mL)=&\left\{(\bW(x),w),w=\alpha (\bsb{n}\cdot \bU_d)\in \mR,\right.\nonumber\\
&\left.\bV_0 = \mY \bU_0,\, (\mDelta\bU'_d)_\perp=0=(\bV_d)_\perp\right\},\notag
\end{align}
where $\bsb{n}$ is a vector projecting on the first transmission line function, i.e.,  $\bsb{n}=
\begin{pmatrix}
1&0&0
\end{pmatrix}^T$, and $\perp$ refers to its orthogonal complement. $\alpha$ is a free parameter which will be optimally set to the value $\alpha_s=C_s/c_\delta=C_c C_J/[c_\delta(C_c + C_J)]$, such that the Hamiltonian will not have mode-mode couplings, see \cite{ParraRodriguez:2018}. For simplicity, we have assumed open boundary conditions (current equals to zero) in the other lines. A suitable inner product is determined by the parameter $\alpha$, and for pairs of elements of the Hilbert space 
\begin{equation}
\langle \mcl{W}_1,\mcl{W}_2\rangle=\int_{\mcl{I}} dx\, \bW_1 \mUpsilon \bW_2+\frac{w_1 w_2}{\alpha},
\end{equation}
where $\mUpsilon$ is defined as previously in (\ref{eq:inner_product_lines_only}). The action of the operator on its elements now reads $\mL\mW=(-\mDelta\bW'', -(\bsb{n}\cdot\mDelta\bU'_d))$. It is easy to check that $\mL$ is self-adjoint with this inner product, which means that its eigenvectors form a basis. The Lagrangian of the system is written as
\begin{eqnarray}
L&=&L_{\text{TG}}+L_{\mY}+\frac{\alpha_c}{2}\left(\bsb{n}\cdot\dot{\bPhi}_d-\frac{\dot{\Phi}_J}{\sqrt{\alpha_\Sigma}}\right)^2+\frac{\alpha_J}{2\alpha_\Sigma}\dot{\Phi}_J^2\nonumber\\
&&+E_J\cos(\varphi_J),
\end{eqnarray}
where we have rescaled the flux and phase variables such that $\alpha_i = C_i/c_\delta$, $C_\Sigma=(C_J+C_c)$, $\varphi_J=2\pi\Phi_J/(\Phi_{q} \sqrt{C_\Sigma})$, and $\Phi_q$ is the flux quantum. 

We must solve the eigenvalue problem  $\mL\mW_{n\epsilon}=\omega_n^2\mW_{n\epsilon}$  to expand the fields in the eigenbasis of the differential operator and rewrite the Lagrangian as 
\begin{align}
L=&\frac{1}{2}\sum_{n,\epsilon} \dot{X}_{n\epsilon}(\dot{X}_{n\epsilon}-\omega_n\msJ_{\epsilon\epsilon'}X_{n\epsilon'})+\frac{\alpha_-}{2}(\dot{X}_{n\epsilon} u_{n \epsilon})^2\nonumber\\
&-\frac{\alpha_c}{\sqrt{\alpha_\Sigma}} (\dot{X}_{n\epsilon} u_{n\epsilon})\dot{\Phi}_J+\frac{\dot{\Phi}_J^2}{2}+E_J\cos(\varphi_J)\nonumber
\end{align}
where $\alpha_-=\alpha_c-\alpha$,  and  $u_{n\epsilon}=\bsb{n}\cdot(\bU_{n\epsilon})_d$. We remark that we have picked the basis for which $i\mt=-i\sigma_y=\msJ^T$, and choose to normalise as $\langle \mW_{n\epsilon},\mW_{m\epsilon'}\rangle=\delta_{nm}\delta_{\epsilon\epsilon'}$. We can write in a more compact notation
\begin{eqnarray}
L=\frac{1}{2}\bsb{\dot{\bar{X}}}^T\mC \left(\bsb{\dot{\bar{X}}}-\msG\bar{\bX}\right)+E_J\cos(\varphi_J),
\end{eqnarray}
where the coordinates' vector is $\bar{\bX}=(\Phi_J,\bX^T)=(\Phi_J, F_{11}, ...,F_{1N}, G_{11},...,G_{1N},...,F_{21},...,G_{21}, ...)^T$, and the matrices are
\begin{align}
\mC&=\begin{pmatrix}
1&-\frac{\alpha_c }{\sqrt{\alpha_\Sigma}}\bu^T\\-\frac{\alpha_c }{\sqrt{\alpha_\Sigma}}\bu & \mone + \alpha_- \bu\bu^T
\end{pmatrix},\nonumber\\
\msG&=\begin{pmatrix}
0&0&0&\\0& \omega_1 \msJ&0&\\0&0& \omega_2 \msJ&\\&&&\ddots
\end{pmatrix},\nonumber
\end{align}
with $\bu=(u_{11}, u_{12}, ..., u_{1(2N)}, u_{21},....)^T$. We perform a Legendre transformation 
$\bar{\bPi}=\partial L/\partial \bsb{\dot{\bar{X}}}=(Q_J,\bPi^T)=(Q_J, \pi_{11},..., \pi_{1N}, p_{11},...,\pi_{1N},...,\pi_{21},...,p_{21},...)^T$, by formally inverting the kinetic matrix $\mC$. 

We fix $\alpha=\alpha_s$, such that the mode-mode coupling disappears, see Eqs. (16-18) 
in \cite{ParraRodriguez:2018}
\begin{equation}
\left.\mC^{-1}\right|_{\alpha=\alpha_s}=\begin{pmatrix}
\frac{\alpha_\Sigma}{\alpha_J}&\frac{\alpha_c}{\sqrt{\alpha_\Sigma}} \bu^T\\\frac{\alpha_c}{\sqrt{\alpha_\Sigma}} \bu& \mone
\end{pmatrix},\nonumber
\end{equation}
where the infinite-length coupling vector has finite norm $|\boldsymbol{u}|^2=1/\alpha=1/\alpha_s$ \cite{Walter:1973,ParraRodriguez:2018}. 

We derive the Hamiltonian 
\begin{align}
H=&\frac{1}{2}\left(\bar{\bPi}+\tfrac{1}{2}\msG\bar{\bX}\right)^T\mC^{-1} \left(\bar{\bPi}+\tfrac{1}{2}\msG\bar{\bX}\right)+E_J\cos(\varphi_J)\nonumber\\
=&\frac{1}{2}\sum_{n,\epsilon}\left( \Pi_{n\epsilon}+\frac{1}{2}\omega_n\msJ_{\epsilon\epsilon'}X_{n\epsilon'}\right)^2+\frac{\alpha_\Sigma}{2\alpha_J}Q_J^2\nonumber\\
&-E_J\cos(\varphi_J)+\gamma Q_J u_{n\epsilon}\left( \Pi_{n\epsilon}+\frac{1}{2}\omega_n\msJ_{\epsilon\epsilon'}X_{n\epsilon'}\right),\nonumber
\end{align}
where $\gamma=\alpha_c/\sqrt{\alpha_\Sigma}$. 

One can apply now the canonical transformation that eliminates the $N$ nondynamical coordinates per frecuency
\begin{eqnarray}
\tilde{F}_{n\lambda}&=&\frac{1}{2}F_{n\lambda}-\frac{1}{\omega_n}P_{n\lambda},\qquad \tilde{\pi}_{n\lambda}=\pi_{n\lambda}+\frac{\omega_n}{2}G_{n\lambda},\nonumber\\
\tilde{G}_{n\lambda}&=&\frac{1}{2}G_{n\lambda}-\frac{1}{\omega_n}\pi_{n\lambda},\qquad
\tilde{p}_{n\lambda}=p_{n\lambda}+\frac{\omega_n}{2}F_{n\lambda}, \nonumber
\end{eqnarray}
and rescale back the Josephson conjugate variables $Q_J\rightarrow Q_J/\sqrt{C_\Sigma}$ and $\varphi_J\rightarrow \varphi_J\sqrt{C_\Sigma}$, to derive the Hamiltonian 
\begin{align}
H=&\frac{1}{2}\sum_{n,\epsilon}\left( \tilde{\pi}_{n\lambda}^2+\omega_n^2\tilde{F}_{n\lambda}^2\right)+\frac{Q_J^2}{2C_J}-E_J\cos(\varphi_J)\nonumber\\
&+\xi Q_J \left(u_{n u \lambda}\tilde{\pi}_{n\lambda}-u_{n v \lambda}\omega_n\tilde{F}_{n\lambda}\right),\nonumber
\end{align}
where $\xi=\gamma/\sqrt{C_\Sigma}=\alpha_c/(\alpha_\Sigma \sqrt{c_\delta})$. We promote the conjugate variables to quantised operators, and define the annihilation and creation pair as $\hat{\tilde{\pi}}_{n\lambda}=\sqrt{\hbar\omega_n/2}(\hat{a}_{n\lambda}+\hat{a}_{n\lambda}^\dag)$, and $\hat{\tilde{F}}_{n\lambda}=i\sqrt{\hbar/(2\omega_n)}(\hat{a}_{n\lambda}-\hat{a}_{n\lambda}^\dag)$. Finally,
\begin{align}
\hat{H}=&\frac{\hat{Q}_J^2}{2C_J}-E_J\cos(\hat{\varphi}_J)+\sum_{n,\lambda}^{\infty,N} \hbar\omega_{n}\, \hat{a}_{n\lambda}^\dag \hat{a}_{n\lambda}\nonumber\\
&+\xi \hat{Q}_J\sum_{n,\lambda}\left[r_{n\lambda}^* \hat{a}_{n\lambda}+r_{n\lambda} \hat{a}_{n\lambda}^\dagger\right]\nonumber
\end{align}
where $r_{n\lambda}=\sqrt{\frac{\hbar\omega_n}{2}}  (u_{n u\lambda}+i u_{n v\lambda})$. This coupling parameter allows us the computation of convergent Lamb-shifts (and effective multi-partite couplings), as shown by   \cite{Malekakhlagh:2017,Gely:2017,ParraRodriguez:2018}, that are proportional to \begin{equation}
\chi\propto \sum_{n,\lambda}|r_n|^2/\omega_n=\hbar/(2\alpha)<\infty, \nonumber
\end{equation}
with convergence due to the eigenfunctions decaying as $u_{n\epsilon}\propto 1/n$ when $\omega_n\rightarrow\infty$.

\bibliographystyle{apsrev4-1}
\bibliography{bibliography}

\end{document}